\documentclass[preprint2]{proto}
\usepackage{natbib}
\usepackage{amsmath}
\usepackage{hyperref}
\bibpunct{(}{)}{;}{a}{,}{,}
\usepackage{times}

\begin{document}

\title{\textbf{\LARGE Dust Evolution in Protoplanetary Disks}}

\author {\textbf{\large Leonardo Testi$^{1,2,3}$, Tilman Birnstiel$^{4}$, Luca Ricci$^{5}$, Sean Andrews$^{4}$, J\"urgen Blum$^{6}$, John Carpenter$^{5}$, Carsten Dominik$^{7}$, Andrea Isella$^{5}$, Antonella Natta$^{2,8}$, Jonathan P. Williams$^{9}$, David J. Wilner$^{4}$}}
\affil{\small\em $^1$European Southern Observatory, Germany}
\affil{\small\em $^2$INAF--Osservatorio Astrofisico di Arcetri, Italy}
\affil{\small\em $^3$Excellence Cluster Universe, Boltzmannstr. 2, D-85748 Garching, Germany}
\affil{\small\em $^4$Harvard-Smithsonian Center for Astrophysics, USA}
\affil{\small\em $^5$California Institute of Technology, USA}
\affil{\small\em $^6$Institut f\"ur Geophysik und extraterrestrische Physik, TU Braunschweig, Germany}
\affil{\small\em $^7$University of Amsterdam, The Netherlands}
\affil{\small\em $^8$Dublin Institute for Advanced Studies, Ireland \&\ INAF--Osservatorio Astrofisico di Arcetri, Italy}
\affil{\small\em $^9$Institute for Astronomy, University of Hawaii, USA}


\begin{abstract}
\baselineskip = 11pt
\leftskip = 0.65in 
\rightskip = 0.65in
\parindent=1pc
{\small
In the core accretion scenario for the formation of planetary rocky cores, the first step toward
planet formation is the growth of dust grains into larger and larger aggregates and eventually
planetesimals. Although dust grains are thought to grow from the submicron sizes typical of
interstellar dust to micron size particles in the dense regions of molecular clouds and cores, the
growth from micron size particles to pebbles and kilometre size bodies must occur in the high
densities reached in the mid-plane of protoplanetary disks. This critical step in the formation of
planetary systems is the last stage of solids evolution that can be observed directly in young
extrasolar systems before the appearance of large planetary-size bodies.

\medskip

Tracing the properties of dust in the disk mid-plane, where the bulk of the material for planet
formation resides, requires sensitive observations at long wavelengths (sub-mm through cm waves). At
these wavelengths, the observed emission can be related to the dust opacity, which in turns depend
on to the grain size distribution. In recent years the upgrade of the existing (sub-)mm arrays, the
start of ALMA Early Science operations and the upgrade of the VLA have  significantly improved the
observational constraints on models of dust evolution in protoplanetary disks. Laboratory
experiments and numerical simulations led to a substantial improvement in the understanding of the
physical processes of grain-grain collisions, which are the foundation for the models of dust
evolution in disks.

\medskip

In this chapter we review the constraints on the physics of grain-grain collisions as they have
emerged from laboratory experiments and numerical computations. We then review the current
theoretical understanding of the global processes governing the evolution of solids in
protoplanetary disks, including dust settling, growth, and radial transport. The predicted
observational signatures of these processes are summarized.

\medskip

We discuss the recent developments in the study of grain growth in molecular cloud cores and in
collapsing envelopes of protostars as these likely provide the initial conditions for the dust in
protoplanetary disks. We then discuss the current observational evidence for the growth of grains in
young protoplanetary disks from millimeter surveys, as well as the very recent evidence of radial
variations of the dust properties in disks. We also include a brief discussion of the constraints on
the small end of the grain size distribution and on dust settling as derived from optical, near-,
and mid-IR observations. The observations are discussed in the context of global dust evolution
models, in particular we focus on the emerging evidence for a very efficient early growth of grains
in disks and the radial distribution of maximum grain sizes as the result of growth barriers in
disks.  We will also highlight the limits of the current models of dust evolution in disks including
the need to slow the radial drift of grains to overcome the migration/fragmentation barrier.
\\~\\~\\~}
\end{abstract}

\section{\textbf{INTRODUCTION}}
\label{Sec:Intro}

In this chapter we will discuss the evolution of dust in protoplanetary disks,
focusing on the processes of grain growth and the observational consequences
of this process. In the standard scenario for planet formation, this is the phase
in which the solids grow from micron-size particles, which are present in the 
molecular cloud cores out of which stars and protoplanetary disks are formed, to 
centimeter size and beyond on the path to become planetesimals. This is the 
last stage of solid growth that is directly observable before the formation of 
large, planetary-size bodies that can be individually observed. As this 
phase of growth is directly observable, it has the potential of setting strong 
constraints on the initial stages of the planet formation process. 
In this review, we focus on the growth of particles on the 
disk mid-plane, as this is where most of the solid mass of the disk is concentrated 
and where planets are expected to form. Grain growth in the disk inner regions
and atmosphere can be effectively investigated in the infrared
and have been extensively reviewed in \citet{2007prpl.conf..767N}, although
at the time of that review the observational evidence and theoretical
understanding of the global dust evolution processes in the disk were limited. 

\medskip

This phase of growth up to  centimeter size particles in 
protoplanetary disks can also be connected to the study of the most 
pristine solids in our own Solar System.
We can directly follow grain growth in the  Solar nebula
through the study of primitive rocky meteorites known as chondrites.
Chondrites are composed mainly of millimeter-sized, spherical chondrules
with an admixture of smaller, irregularly-shaped calcium aluminum inclusions
(CAI) embedded in a fine-grained matrix \citep[e.g.][]{2005ApJ...623..571S}.
Some chondrules may be the splash from colliding planetesimals
but most have properties that are consistent with being flash-heated
agglomerates of the micron-sized silicate dust grains in the matrix.
CAIs are the first solids to condense from a hot, low pressure gas
of solar composition.  Their ages provide the zero point for
cosmochemical timescales and the astronomical age of the Sun, 4.567 Gyr.
Relative ages can be tracked by the decay of short-lived radionucleides and
allow the formation timescales of dust agglomerates and planetesimals to be
placed in context with astronomical observations \citep{2011AREPS..39..351D}.
Recent measurements of  absolute ages of the chondrules 
found that these were formed over a period that span from CAI formation to a few Myr
beyond \citep{2012Sci...338..651C}.  
As chondrule dating refers to the time when the dust agglomerate
was melted, this new result tells us  either that dust grains rapidly grew up 
to millimeter sizes in the early solar nebula and were melted progressively 
over time while agglomeration into 
larger bodies continued over the lifetime of the protoplanetary disk; or that
the growth of these millimeter and centimeter size agglomerates from smaller 
particles continued over a period of several million years. In any case,
the process of assembling CAIs and chondrules into larger bodies had
to occur throughout the disk lifetime (see also the chapter by
\textit{Johansen et al.}).

\medskip

In this review we will thus concentrate on the most recent developments 
of the theoretical models for grain evolution in disks and the constraints
on the grain-grain collisions from laboratory experiments as well as on the 
new wealth of data at (sub-)millimeter and centimeter wavelengths that is 
becoming available. The goal will be to give an overview of the current
astronomical constraints on the process and timescale of grain growth 
in disks.

\medskip

The processes and observations that we discuss in this chapter are expected
to occur in protoplanetary disks and both the theoretical models and the interpretation
of the observations rely on assumptions on the disk structure and its evolution.
The structure and evolution of protoplanetary disks during the pre-main 
sequence phase of stellar evolution has been discussed extensively in 
recent reviews and books 
\citep[e.g.][]{2007prpl.conf..555D,2009apsf.book.....H,2010apf..book.....A,2011ARA&A..49...67W}.
Throughout this chapter, our discussion will assume a flared, irradiated disk
structure in hydrostatic equilibrium with a constant gas to dust ratio of
100 by mass, unless explicitly stated otherwise. While the detailed disk 
structure and its evolution under the effects of viscous accretion, chemical
evolution, photoevaporation and planet-disk interaction are actively investigated
in detail (see e.g. the chapters by \textit{Alexander et al.}, 
\textit{Audard et al.}, \textit{Dutrey et al.},
and \textit{Turner et al.}), this assumed disk structure is 
an adequate representation of the early phases of disk evolution and a 
good reference for the processes of dust evolution. 

\begin{figure*}[thb]
\centering
\includegraphics[width=18cm]{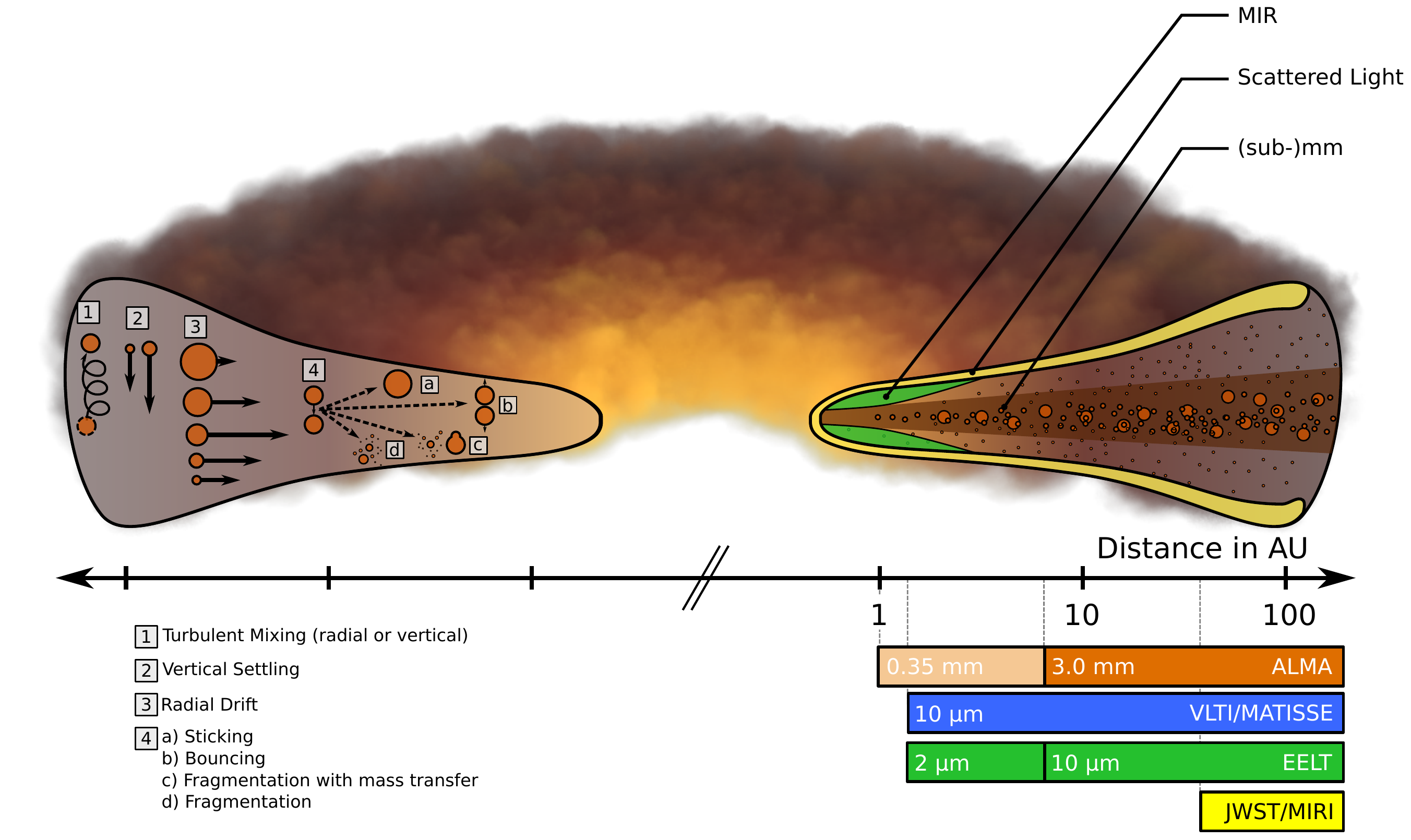}
\caption{Illustration of the structure, grain evolution processes and observational
constraints for protoplanetary disks. On the left side we show the main grain 
transport and collision mechanism properties. The different lengths of the arrows
illustrate the different velocities of the different grains. On the right hand
side, we show the areas of the disk that can be probed by the various techniques.
The axis shows the logarithmic radial distance from the central star.
The horizontal bars show the highest angular resolutions (left edge of the bars)
that can be achieved with a set of upcoming facilities and instruments for 
at the typical distance of the nearest star forming regions.
}
\label{fig:sketch}
\end{figure*}

Figure~\ref{fig:sketch} shows a sketch of a protoplanetary disk where we 
illustrate pictorially the physical processes involved in grain growth on the left side. On the right side 
we illustrate the regions probed by the various observational techniques and the 
angular resolutions offered by the forthcoming generation of facilities and
instruments in the infrared and submillimetre regions.

\medskip

In \S~\ref{Sec:Processes}, we introduce the main concepts at the basis of 
the transport, dynamics and evolution of solid particles in disks. 
First we introduce the interactions between the solids and gas and the 
complex dynamics of solids in disks, then we describe the basic processes
of the growth of solids towards larger aggregates. In \S~\ref{Sec:Lab}
we discuss the constraint on the outcomes of grain-grain collisions from 
laboratory experiments and numerical computations. These provide the basic
constraints to the global evolution models of solids in disks, which we describe
in \S~\ref{Sec:Theo}. In \S~\ref{Sec:Opacities} we discuss the effects of
growth on the dust opacities and the observational consequences. The possible
recent evidence for grain evolution at and before the disk formation epoch is 
discussed in \S~\ref{Sec:pre-disc}. The observational constraints for grain growth 
at infrared wavelengths are described in \S~\ref{Sec:scattir}.
We discuss grain properties 
in the disk mid-plane from (sub-)millimeter observations in
\S~\ref{Sec:Obs}, beginning with the methodology in \S~\ref{Sec:Method},
then the results of low resolution, multi-wavelength continuum surveys in \S~\ref{Sec:sub-mm_phot},
ending with the most recent resolved studies in \S~\ref{Sec:Resolved}.

%
%
\section{\textbf{DUST TRANSPORT AND GROWTH PROCESSES IN DISKS}}
\label{Sec:Processes}

\subsection{\bf Dust transport processes in disks}\label{theo:transport}

\subsubsection{Drag forces}\label{theo:transport:drag_force}
Dust particles embedded in a gaseous protoplanetary disks are not orbiting freely, but feel a
friction when moving with respect to the gas. The force exerted on them depends not only on the
relative motion between gas and dust, but also on the particle size: small particles that are
observable at up to cm wavelength can quite safely be assumed to be smaller than the mean free
path of the gas molecules and are thus in the Epstein regime. If the particles are larger than about
the mean free path of the gas molecules, a flow structure develops around the dust particle and the
drag force is said to be in the Stokes drag regime \citep{1972fpp..conf..211W,1977MNRAS.180...57W}.
Large particles in the inner few AU of the disk could be in this regime, and the transition into the
Stokes drag regime might be important for trapping of dust particles and the formation of
planetesimals \citep[e.g.][]{2010A&A...513A..79B,Laibe:2012p18591,2012ApJ...752..106O}.
An often used quantity is the stopping time, or friction time, which is the characteristic time
scale for the acceleration or deceleration of the dust particles $\tau_\mathrm{s} = m\,v/F$, where
$m$ and $v$ are the particle mass and velocity, and F is the drag force. Even more useful is the
concept of the Stokes number, which in this context is defined as
\begin{equation}
\mathrm{St} = \Omega_\mathrm{K}\,\tau_\mathrm{s},
\label{eq:theo:stokes_def}
\end{equation}
a dimensionless number, which relates the stopping time to the orbital period $\Omega_\mathrm{K}$.
The concept of the Stokes number is useful because particles of different shapes, sizes, or
composition, or in a different environment have identical aerodynamical behavior if they have the same
Stokes number.

\subsubsection{Radial drift}\label{theo:transport:drift}
The simple concept of drag force leads to important implications, the first of which, \textit{radial
drift}, was realized by \cite{Whipple:1972p4621}, \cite{1976PThPh..56.1756A}, and by
\citet{1977MNRAS.180...57W}: an orbiting parcel of gas is in a force balance between gravitational,
centrifugal, and pressure forces. The pressure gradient is generally pointing outward because
densities and temperatures are higher in the inner disk. This additional pressure support results is
a slightly sub-Keplerian orbital velocity for the gas. In contrast, a freely orbiting dust particle
feels only centrifugal forces and gravity, and should therefore be in a Keplerian orbit. This slight
velocity difference between gas and a free floating dust particle thus causes an efficient
deceleration of the dust particle, once embedded in the gaseous disk. Consequently, the particle
looses angular momentum and spirals towards smaller radii. This inward drift velocity is only a
small fraction of the total orbital velocity (a few per mille), but, for St$\sim$1 particles, it
still leads to an inward drift speed of the order of 50 m~s$^{-1}$. It also means that particles of
different sizes acquire very different radial velocities and also that at any given radius, dust of
the right size may be quickly moving towards the central star.

\medskip

\citet{1986Icar...67..375N} investigated the equations of motion for arbitrary 
gas to dust ratios. For a value of 100 by mass, gas is dynamically 
dominating and the classical results of \citet{1977MNRAS.180...57W} are 
recovered, however for a decreasing ratio, the drag that the dust exerts
on the gas becomes more important, and eventually may be reversed: dust 
would not drift inward and gas would be pushed outward instead. Much lower gas-to-dust ratios, approaching 
unity, however, also lead to other effects such as the streaming
instability \citep{Youdin:2005p11585} or self-induced stirring (for details, 
see the chapter by \textit{Johansen et al.}).

\medskip 

The radial drift process does not necessarily mean that all particles are
falling into the star, as dust may pile up at some specific locations 
in the innermost regions of the disk where the gas to dust ratio will 
become very low
\citep[this process has actually been suggested as an explanation for the abundance of rocky exoplanets close to the host star][]{2014ApJ...780...53C},
however a significant fraction of large grains need to be kept in the outer disk
for long timescales, otherwise it would result in a stark contrast to the observed 
population of mm-sized grains in the outer disk (see Sect.~\ref{Sec:Obs}).

\medskip 

In the next paragraph we will discuss some processes that may oppose
and slow down the radial drift.

\subsubsection{Dust Trapping}\label{theo:transport:trapping}
Many works have addressed details of radial drift \citep[e.g.]
{1972fpp..conf..211W,1977MNRAS.180...57W,2002ApJ...580..494Y,Takeuchi:2002p3167,Brauer:2007p232},
however the main conclusions remain: unless the gas to dust ratio is very low, or disks are much
more massive than seems reasonable, \textit{observable} particles should drift to the inner regions
within only a small fraction of the life times of protoplanetary disks. \textit{Observable} here
means detectable with current (sub-)millimeter observatories, i.e., mainly dust
in the outer regions (beyond $\sim$20~AU) of the disk at current resolutions. 
The only other way to stop particles from spiraling inward is to locally 
reverse the pressure gradient, as can be seen from the drift velocity
\citep{1986Icar...67..375N}
\begin{equation}
u_\mathrm{drift} = \frac{1}{\mathrm{St}+\mathrm{St}^{-1}\,(1+\epsilon)^2}
\frac{c_\mathrm{s}^2}{V_\mathrm{K}}\,\frac{\partial \ln P}{\partial \ln r},
\label{eq:theo:v_drift}
\end{equation}
where $P$ is the gas pressure, $c_\mathrm{s} = \sqrt{k_\mathrm{B}\,T/\mu\,m_\mathrm{p}}$ the
isothermal sound speed, $V_\mathrm{K}$ the Keplerian velocity, and $\epsilon$ the dust-to-gas ratio.
If the pressure gradient is zero or positive, there is no radial drift or 
particles drift outward.

\medskip

This basic mechanism of dust drifting to regions of higher pressure has been explored in different
settings: \citet{1995A&A...295L...1B} showed that anticyclonic vortices represent a high-pressure
region, which can accumulate dust particles \citep[see also
][]{1997Icar..128..213K,Fromang:2005p20823}. The dust drift mechanism also works efficiently in the
azimuthal direction if there exist regions of azimuthal over-pressure
\citep[e.g.][]{2013A&A...550L...8B}, however it should be stressed, that the mechanism relies on
relative motion between gas and dust: as an example of this exception, an over-density caused by an
eccentric disk, does not trap dust particles, as shown  by \citet{Hsieh:2012p21441} and
\citet{Ataiee:2013p20827}. For an eccentric disk, the over-density, and thus the pressure maximum,
arises due to the non-constant azimuthal velocity along the elliptic orbits, but the same holds for
the dust, i.e., the velocity difference between dust and the partially pressure supported gas
remain, and consequently, the radial drift mechanism is still active, but the dust is not
concentrated azimuthally with respect to the gas.

%

\subsubsection{Radial mixing and meridional flows}\label{theo:transport:radial_mixing}
The drift motion towards higher pressure is not the only mode of transport 
of dust, there are two additional ones: mixing and advection. These are 
interesting as they may oppose radial drift under certain circumstances 
and may provide an explanation for mixing processed dust outward in the disk.
Outward mixing processes, including winds \citep[e.g.][]{Shu:1994p20679,Shu:2001p20834},
have been invoked to explain the presence of crystalline
solids in disks \citep{2004A&A...415.1177K,Ciesla:2009p10132} and comets 
in our own Solar System \citep{BockeleeMorvan:2002p20828}. Evidence for 
dust processing in the inner regions of disks and subsequent radial 
transport has been recently observed directly by Spitzer 
\citep{2009Natur.459..224A,2012ApJ...744..118J}.
In this review we will not discuss these topics that have been extensively 
covered elsewhere 
\citep[e.g.][]{2007prpl.conf..767N,2010ARA&A..48..205D,2011ARA&A..49...67W},
below we just provide a brief account of the radial transport processes.

Advection is a result of the viscous evolution of the
gas which essentially drags dust particles along with the radial gas 
velocity, as long as they are efficiently coupled to the gas. This drag 
velocity was derived, for example in \citet{Takeuchi:2002p3167} as
\begin{equation}
u_\mathrm{r,drag} = \frac{u_\mathrm{r,gas}}{1+\mathrm{St}^2}, 
\end{equation}
which states that only small particles (St$<1$) are effectively coupled with 
the gas.
It was first shown by \citet{1984SvA....28...50U}, that the gas velocity of a region with net inward
motion can be positive at the mid-plane. This flow pattern, called \textit{meridional flow},
allows for outward transport of dust grains. 
This process is typically much weaker than radial drift and thus
unable to explain  millimeter sized particles in the outer disk. 
In addition,  large-scale circulation pattern of meridional flows, while present 
in viscous disk simulations \citep[][]{Kley:1992p7134,Rozyczka:1994p20829}, 
are not reproduced in MRI turbulent simulations
\citep{2011A&A...534A.107F}. 

\medskip

In addition to this systematic motion, the gas is also thought to be turbulent (see also the
chapters by \textit{Dutrey et al.} and by \textit{Turner et al.}), and the dust is therefore mixed
by this turbulent stirring as well. The fact that the dust is not necessarily perfectly coupled to
the gas motion leads to some complications \citep[see,][]{Youdin:2007p2021}. The ratio between gas
viscosity $\nu_\mathrm{gas}$ and gas diffusivity $D_\mathrm{gas}$ is called the Schmidt number. It
is usually assumed that the gas diffusivity equals the gas viscosity, in which case the Schmidt
number for the dust is defined as
\begin{equation}
\mathrm{Sc} = \frac{D_\mathrm{gas}}{D_\mathrm{dust}}.
\end{equation}
\citet{Youdin:2007p2021} included effects of orbital dynamics, which were neglected in previous
studies of \citet{Cuzzi:1993p15730} and \citet{Schrapler:2004p2394}, and showed that the Schmidt
number can be approximated by $\mathrm{Sc} \simeq 1+\mathrm{St}^2$.
It is currently believed that none of these radial transport mechanisms can overcome the radial
drift induced by the pressure gradient because the vertically-integrated net transport velocity is
dominated by the much stronger radial drift velocity \citep[e.g.][]{Jacquet:2013p21188}.

\subsubsection{Vertical mixing \& settling}\label{theo:transport:vertical_settling_mixing}
A dust particle elevated above the gas mid-plane, orbiting at the local Keplerian velocity, would
vertically oscillate due to its orbital motion, if it were not for the gas drag force, which damps
motion relative to the gas flow. Following the concepts discussed in the previous sections,
particles with Stokes number smaller than unity are thus damped effectively within one orbit. Larger
particles will experience damped vertical oscillations. The terminal velocity of small particles can
be calculated by equating the vertical component of the gravitational acceleration and the
deceleration by drag forces. The resulting settling velocity for $\mathrm{St}<1$, $v_\mathrm{sett} =
\mathrm{St} \, \Omega_\mathrm{K} \,z$, increases with particle size and height above the mid-plane.

\medskip

Vertical settling was already  the focus of the earliest models of 
planetesimal formation, such as \citet{Safronov:1969p11177}, 
\citet{Goldreich:1973p11184}, \citet{1980Icar...44..172W}, and
\citet{1986Icar...67..375N}, and has remained an
active topic of debate
\citep[e.g.][]{Cuzzi:1993p15730,Johansen:2005p8425,Carballido:2006p18250,Bai:2010p15702}.
The main question, however, is usually not the effectiveness of settling motion itself, but the
effectiveness of the opposing mechanism: turbulent mixing. \citet{1995Icar..114..237D} calculated
the vertical structure of the dust disk, by solving for an equilibrium between settling and mixing
effects. The resulting scale height of the dust concentration in an isothermal disk with scale
height $H_\mathrm{p} = c_\mathrm{s}/\Omega_\mathrm{K}$ then becomes $H_\mathrm{dust} = H_\mathrm{p}
/ \sqrt{1+\mathrm{St} / \alpha_\mathrm{t}}$, where we used the canonical turbulence prescription with parameter
$\alpha_\mathrm{t}$ \citep{Shakura:1973p4854}. Detailed MHD models of MRI turbulent disks of
\citet{2009A&A...496..597F} show, that the simple result from \citet{1995Icar..114..237D} is a good
approximation only near the mid-plane. Above roughly one scale height, the vertical variations in
the strength of the turbulence and other quantities cause strong variations from the result of
\citet{1995Icar..114..237D}.

\medskip

The effects that dust settling has on the observational appearance of disks have been derived for
example by \citet{Chiang:2001p7463}, \citet{DAlessio:2001p4130}, or \citet{2004A&A...421.1075D}.
While the settling process could lead to the rapid growth of particles and the formation of a thin
mid-plane layer of large pebbles, it is expected that small dust is replenished in the disk upper
layers, e.g. by small fragments from shattering collisions between dust grains
\citep[e.g.][]{2005A&A...434..971D,2009A&A...503L...5B,Zsom:2011p21105} or by continuous infall
\citep[e.g.][]{Mizuno:1988p20730,2008A&A...491..663D}.


\subsection{\bf Grain growth processes}\label{theo:growth}

The transport mechanisms  discussed in the previous sections
all lead to large differential vertical and radial motion of dust particles.
These in turn imply frequent grain-grain collisions, potentially leading to 
growth. The two main ingredients for a model of dust growth are the 
collision frequency and the collision outcome. The growth process is modeled
as the result of primordial dust particles, referred to as {\it monomers}, that 
can stick together to form larger {\it aggregates}.
The latter depends on many different parameters, such as the composition 
(e.g. fraction of icy, silicate, or carbonaceous particles), the monomer 
size distribution, the structure (i.e., compact, porous, fractal grains, 
layers, etc.), the impact parameter and impact velocity. We will
first discuss here the expected ranges of impact velocities and then, in
\S~\ref{Sec:Lab}, some of the recent laboratory constraints on the 
collision outcomes.

\medskip

The final ingredient, the collision frequency, depends on the one hand on 
the cross section of the particles and their number density, which are 
results of the modeling itself, and on the other hand on the relative 
velocity of grains. We will describe these in the context of the global dust
evolution models in \S~\ref{Sec:Theo}.

\subsubsection{Relative velocities}\label{theo:growth:relative_velocities}
%
%
The relative velocities between particles in disks under the combined effects of settling and radial
drift can directly be derived from the terminal dust velocities of \citet{1986Icar...67..375N} for
two different-sized particles \citep[e.g.][]{1993prpl.conf.1031W,2008A&A...480..859B} and many
others. As an example, the upper panel of Fig.~\ref{fig:lab} shows the expected relative velocities
between grains of different sizes as computed for 1~AU by \citet{1993prpl.conf.1031W}.
Both radial drift and vertical settling velocities peak at a Stokes number of unity.
Relative motions increase 
with the size difference of the particles because particles with the same Stokes
number have the same systematic velocities. The
maximum relative velocity via radial drift is
\begin{equation}
\Delta v_\mathrm{drift} = \frac{c_\mathrm{s}^2}{V_\mathrm{K}}\,\frac{\partial \ln{P}}{\partial \ln
r}.
\label{eq:theo:dv_drift}
\end{equation}

\medskip

Relative azimuthal velocities are small for particles of Stokes number less than unity, approach a
constant, high value for $\mathrm{St}>1$, and also increase with the Stokes number difference.

\medskip

In addition to these systematic motions, random motions also induce relative velocities, even
between particles of the same Stokes number. Brownian motion of the particles is negligible for
large particles, but it is the dominating source of relative motion for small particles, roughly
sub-$\mu$m sizes.

Much more complicated than the relative motion discussed above and a topic of current research is
turbulent motion. The most frequently used formalism of this problem was introduced by
\citet{1980A&A....85..316V} and \citet{Markiewicz:1991p9934}, and recently, closed-form expressions
were derived by \citet{Cuzzi:2003p12258} and \citet{2007A&A...466..413O}. Their results show that,
similar to radial drift and vertical settling, turbulent relative velocities increase with the
Stokes number difference between the colliding particles and generally increase with the Stokes
number until a $\mathrm{St}=1$. Beyond, it drops off again, but slower than for example relative
velocities induced by radial drift or vertical settling.

The maximum turbulent velocity according to \citet{2007A&A...466..413O} is
\begin{equation}
\Delta v_\mathrm{turb} = c_\mathrm{s}\, ( 9\alpha_\mathrm{t}/2 )^{1/2}
\label{eq:theo:dv_turb}
\end{equation} 
and a factor of $\sqrt{3}$ smaller for collisions between equal-sized particles. 
For typical parameter choices ($\partial\ln P/\partial \ln r$ = 2.75), Eqs.~\ref{eq:theo:dv_drift}
and \ref{eq:theo:dv_turb} imply that the turbulence parameter $\alpha_\mathrm{t}$ has to be only
larger than about $2 (H_\mathrm{p}/r)^2$ to be the dominant source of relative velocity between
grains. The dominant contributions to the relative velocities are also shown in the top panel of
Fig.~\ref{fig:theo1}. For planetesimals, also the gravitational torques of the turbulent gas play a
role (not shown in Fig. ~\ref{fig:theo1}, but see chapter by \textit{Johansen  et al.})

\medskip

The works mentioned above generally treat only the r.m.s. velocity between dust grains, but
the distribution of the collision velocity can also be important. Recently, some numerical works have
tested the analytically derived collision velocities and started to derive distributions of
collision velocities, e.g.
\citet{2010MNRAS.405.2339C,Pan:2010p20918,Hubbard:2012p20911,2013ApJ...776...12P}. We will come back
to the treatment of grain velocities in the context of global disk models in
\S~\ref{theo:modeling:zero_dimensional}.

\subsubsection{Condensation}\label{theo:growth:condensation}
A different physical process for growing (destroying) dust grains without grain collisions is
condensation (evaporation) of material from the gas phase or the sublimation of mantles and solids.
Dust growth via this mechanisms was mentioned already in early works \citep{Goldreich:1973p11184}.
It is commonly discussed in the context of dust formation and evolution in the interstellar medium
\citep[e.g.][]{Zhukovska:2008p16575,Hirashita:2011p16524}, but also in the context of the
``condensation sequence'' \citep[e.g.][]{Lodders:2003p6651}. The problem with growing large grains
via condensation in a protoplanetary disk is twofold: firstly, there is usually not enough material
in the gas phase to grow a macroscopic dust/ice mantle on every microscopic grain, secondly,
accretion is a surface effect and the dust surface area is strongly dominated by the smallest
grains. Condensation will therefore happen preferentially on the smallest grains, until all
condensable material is consumed. Growth of large grains by condensation in protoplanetary disks can
therefore only happen if there is some continuous source of condensable material, (e.g. near an
evaporation zone) and if some mechanism is able to preferentially transport the condensing material
onto the largest particles. Some recent work by \citet{Ros:2013p20760} simulated growth of decimeter
sized particles consisting entirely of ice. However this work relies on the absence of both dust and
radial drift, i.e. dust nuclei are not left over after evaporation and vapor thus recondenses on
another ice particle instead. Whether large particles can be formed by condensation under realistic
disk conditions remains a topic for future research.

%
%
\section{\textbf{CONSTRAINTS ON THE PHYSICAL PROCESSES FROM LABORATORY EXPERIMENTS}}
\label{Sec:Lab}


The constraints on the collision outcomes as a function 
of particle size, composition, shape and relative velocities are an 
essential input to the models of grain evolution in disks. It is not 
easy to explore the full parameter space for the conditions present in
protoplanetary disks, here we will
cover the emerging trends from laboratory experiments and numerical
computations of grain-grain (or aggregate-aggregate) low-velocity collisions 
\citep[for recent reviews, see also][]{Dominik:2007p1420,2008ARA&A..46...21B}.
Dust grains in protoplanetary disks are subject to 
relatively low-velocity mutual collisions, which can have a very wide 
spectrum of results ranging from the complete fragmentation of the two 
particles leading to a swarm of smaller fragments to the formation of a single 
larger particle containing the total mass of the system. 
The experimental techniques of investigating dust-particle and dust-aggregate 
collisions have considerably improved over the past decades so that 
reliable conclusions on the dust evolution in disks can be drawn.

\medskip

The main questions that we will address are the following:

\begin{enumerate}
  \item Why do dust particles or dust aggregates grow?
  \item What are the structures of growing dust particles?
  \item How fast do dust aggregates grow?
  \item What is the maximum size that dust aggregates can reach?
  \item Is the formation of larger aggregates by direct sticking collisions possible?
\end{enumerate}

\medskip

We will review the status of research on the above points one by one. In 
Fig.~\ref{fig:lab} we provide a schematic representation of the various
processes discussed below.

\subsection{\bf Why do dust particles or dust aggregates grow?}

The cause of dust growth is probably the easiest to answer. 
It is generally assumed that in the dense regions of protoplanetary
disks not too close to the central star, electrostatic charges and magnetic 
materials do not play a dominant role in the interaction between grains,
however, this is currently debated and several studies have investigated 
their effects on grain evolution in disks \citep{2002Icar..157..173D,2012ApJ...744....8M,2011ApJ...731...96O}.
In the absence of these effects,
dust grains in contact still possess an adhesive (van der Waals) force  
\citep{1999PhRvL..83.3328H,2011Icar..214..717G}. If the dissipation of 
kinetic energy during the collision is strong enough, hysteresis effects 
at the contact point can lead to the sticking of the dust grains 
\citep{1993ApJ...407..806C,1997ApJ...480..647D}. This direct hit-and-stick coagulation 
process has been experimentally investigated by \citet{2000ApJ...533..472P}.
They found that $\sim 1\rm \mu m$ sized silica grains can stick to one 
another if they collide at less than $\sim 1 \rm m~s^{-1}$. Smaller 
grains can stick at higher velocities, whereas larger grains possess 
a lower threshold for sticking. This threshold velocity, below which 
sticking is dominant, is possibly strongly affected by the grain material. 
Recent investigations have shown that micrometer-sized water-ice 
particles seem to stick up to about 30 m\,s$^{-1}$ collision velocity, 
in agreement with the much higher surface energy 
of water ice \citep{2011Icar..214..717G}. Such high sticking thresholds 
have been used before in numerical simulations of ice-aggregate growth 
\citep[see, e.g.][and the chapter by \textit{Johansen et al.}]{2009ApJ...702.1490W}. 
That larger particles have a lower sticking threshold is 
in principle still true for dust aggregates, although the details of 
the hit-and-stick process for soft-matter (or granular) particles are 
somewhat different. Based on laboratory experiments on collisions among 
(sub-)mm-sized high-porosity dust aggregates (consisting of $\sim 1\rm 
\mu m$ sized silica grains), \citet{2013Icar..225...75K} derived a 
mass-dependent threshold velocity of the form $v \propto m^{-3/4}$. 
However, the threshold velocity for sticking of dust aggregates depends 
on many parameters, e.g. on grain size and material (sub-$\rm \mu m$ 
vs $\rm \mu m$, silicates vs ice), aggregate morphology (fractal, fluffy, 
compact, hierarchical), or mass ratio (through reduced mass).

\subsection{\bf What are the structures of growing dust particles?}

As long as the collision velocity is well below the threshold for 
sticking (see above), a mutual collision is in the so-called 
hit-and-stick regime, in which the projectile particle sticks to the 
target aggregate at the point of first contact. Numerical simulations 
as well as experiments have shown that in this regime, the dust 
aggregates grow to extremely fluffy structures, which have a fractal 
dimension $D$ of well below 2. This is particularly true as long as 
Brownian motion dominates the collisions, which lead to fractal clusters 
with $D \approx 1.5$ 
\citep{2000PhRvL..85.2426B,2004PhRvL..93b1103K,2006Icar..182..274P}. 
This is typically the case during the very first stages of growth, when the 
dust aggregates are not much larger than tens of microns.
With increasing dust-aggregate mass, 
the hit-and-stick regime is being replaced by the compaction stage 
\citep{2000Icar..143..138B,1997ApJ...480..647D} so that the fractal 
dimension increases (whether to values $D=3$ or slightly below is still 
under debate) and dust aggregates are better described by their density 
or porosity. Typical volume filling factors of 0.05-0.50 are being 
expected for most of the mass range \citep{2010A&A...513A..57Z}, although 
\citet{2008ApJ...677.1296W} derived in their numerical simulations growing 
aggregates with D=2.5 (and correspondingly extremely small volume 
filling factors) when 0.1 $\mu$m-sized ice/dust monomers were considered. 
For monomer-size distributions according to the one derived by Mathis
et al. (1977, MRN-type distribution) for interstellar dust particles,
the bulk of the mass of an aggregate is dominated by the largest
monomers, whereas the number of particles and, thus, the contacts
between monomer grains, are determined by the smallest particles.
Ormel et al. (2009) show that such an aggregate has the same strength
as an aggregates of single-size monomers of 0.056~$\mu$m.
These results may also hold for 
aggregates consisting of ice and silicate 
particles. Here, the weaker-bound silicate grains might determine 
the strength and growth properties of the ice-dust bodies.

\begin{figure*}[ht]
\includegraphics[width=7.8cm]{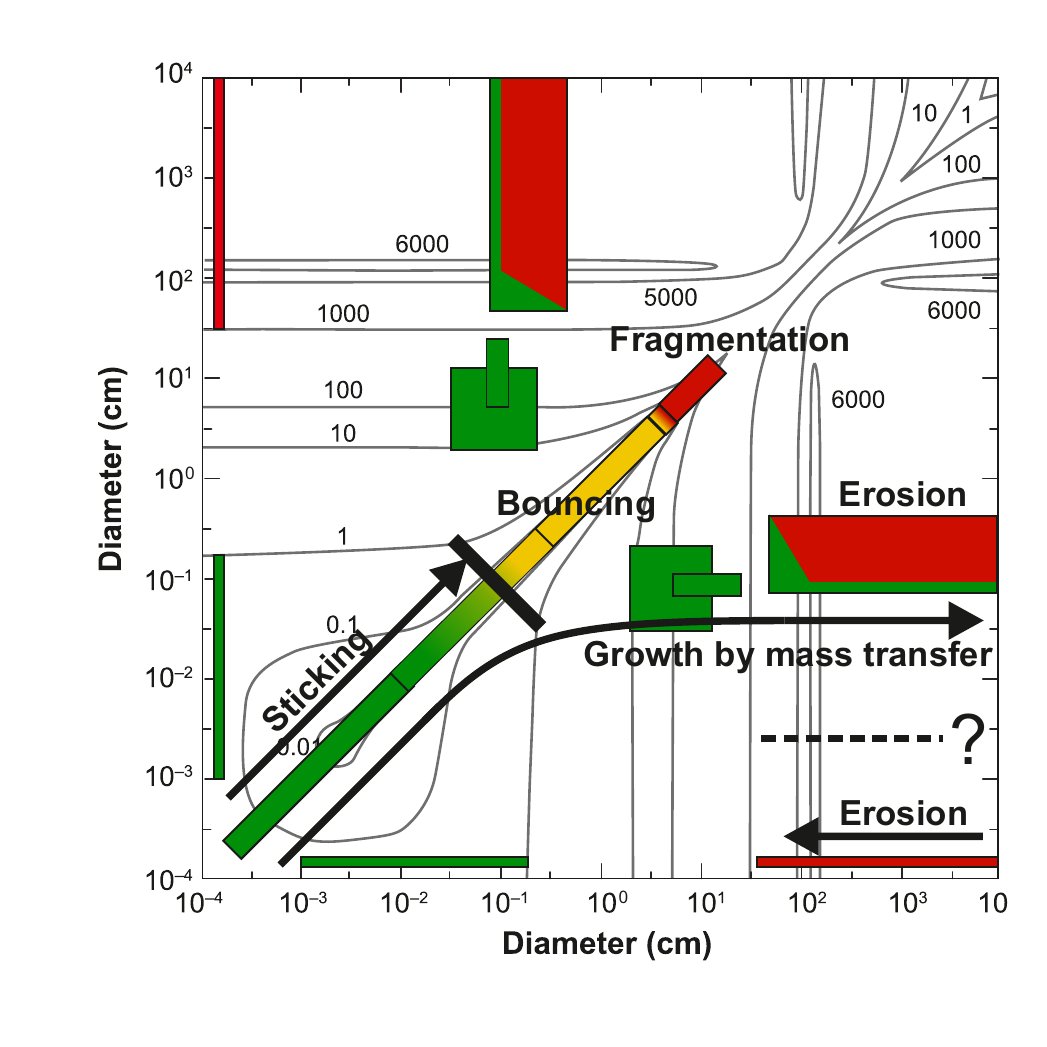}
\includegraphics[width=8.1cm]{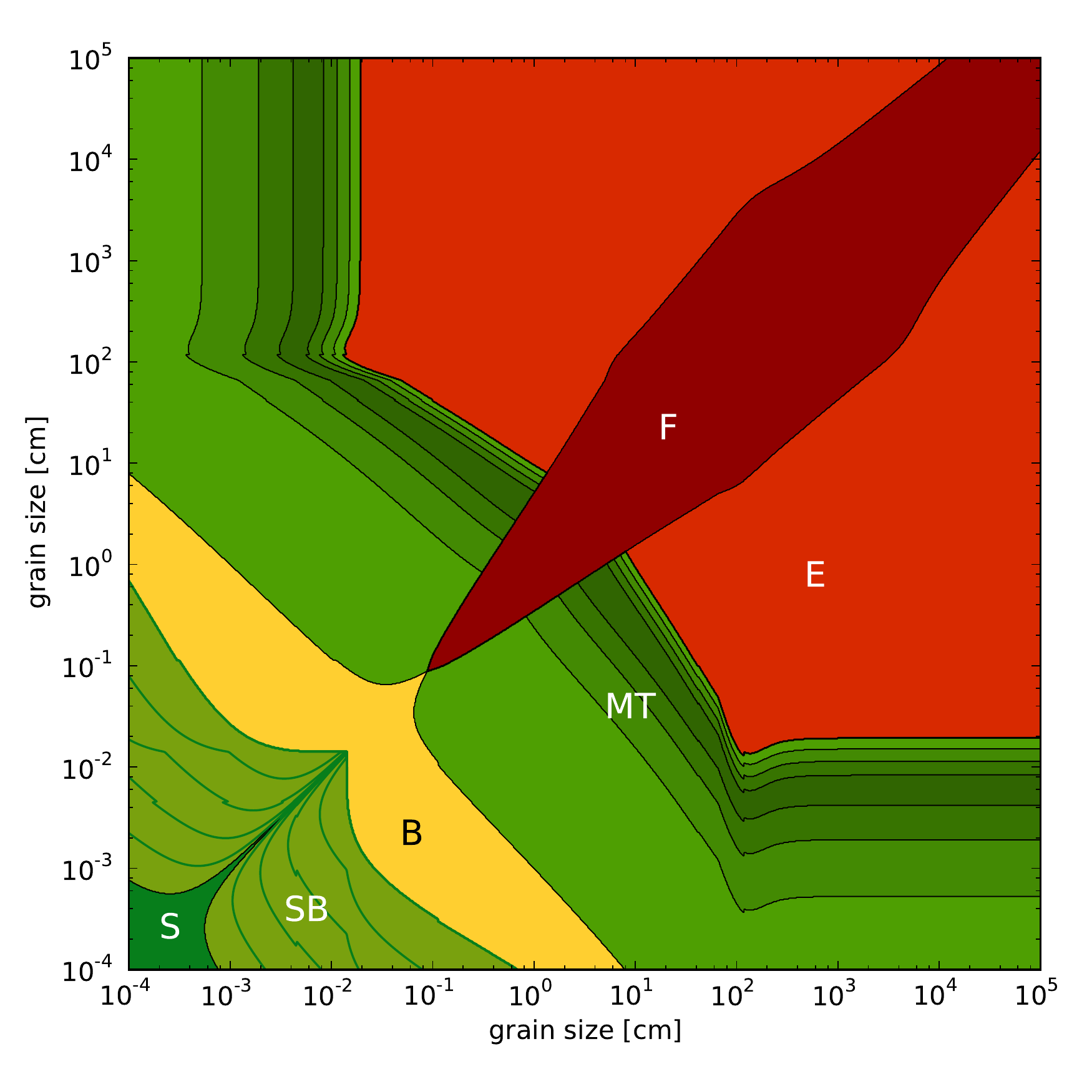}
\caption{Schematic representation of the outcomes of dust collisions 
in protoplanetary disks. Left panel: the background plot  
shows the collision velocities 
(in units of cm/s) between two non-fractal dust agglomerates with 
sizes indicated on the axes in a minimum-mass solar nebula (MMSN) 
model at 1~AU \citet{1993prpl.conf.1031W}. 
The green, yellow, and red boxes denote the explored parameter space
and results 
of laboratory experiments. Here, green represents 
sticking or mass transfer, yellow bouncing, and red fragmentation or erosion. 
The arrow denoted "Sticking" indicates the direct growth of mm-sized 
dust aggregates as found in the simulations by \citet{2010A&A...513A..57Z}.
Further growth is prevented by bouncing. A possible direct path to the 
formation of planetesimals is indicated by the arrow "Growth by mass 
transfer". 
Right panel: the collisions outcomes parameter space as used by numerical 
models of dust evolution in protoplanetary disks by \citet{2012A&A...540A..73W}, as derived 
interpolating the results of the laboratory experiments across the 
entire parameter space. Note that these collisional outcomes 
refer only to collisions between bare silicates grains.
}
\label{fig:lab}
\end{figure*}

\subsection{\bf How fast do dust aggregates grow?}

Both experiments and numerical computations show that the 
first stage of growth is characterized by sticking.
In this early phase their growth rate is simply given 
by the product of the number density of available collision partners, 
their mutual collision cross section, and their relative velocity.
Under the assumption of non-fractal growth, the latter two are easily 
determined, while the first is a result of the growth process 
\citep[see][for a detailed description of dust-growth processes]{2006AdPhy..55..881B}. 
The initial growth is also connected with a compaction of the aggregates
as larger and larger grains collide and stick absorbing part of the kinetic
energy rearranging their internal structure. This phase is followed 
by the so called ``bouncing barrier'', when compact grains will not
easily grow through the hit-and-stick mechanism  \citep{2010A&A...513A..57Z}.
The detailed outcome of the full process of grain growth will be
discussed in the context of the global models in \S~\ref{Sec:Theo},
numerical experiments show that, due to compaction processes, the 
dust aggregates start very fluffy but are ultimately relatively compact.
It is thus realistic to constrain the collision of particles beyond the 
``bouncing barrier''
with the results of laboratory experiments
based on compact aggregates.

\subsection{\bf What is the maximum size that dust aggregates can reach?}

Although many uncertainties about the collision properties of growing
protoplanetary dust aggregates exist 
\citep[see, e.g., the discussion in][]{2013Icar..225...75K}, 
it seems to be evident from laboratory experiments that direct 
sticking between dust aggregates in mutual collisions is limited 
(a possible mass-velocity threshold relation has been mentioned above). 
Thus, there is a maximum dust-aggregate size that can be formed 
in direct sticking collisions. 
The limiting factor for further growth is the bouncing and compaction 
of the dust aggregates, which has been observed in many laboratory 
experiments (see Fig.~\ref{fig:lab}). However, another physical process
which has been found in the laboratory is a possible loophole out of 
this ``bouncing barrier''. Above a certain velocity threshold, which 
is typically close to the fragmentation barrier ($\sim 1 ~\rm m~s^{-1}$), 
collisions between small and large dust aggregates can lead to a 
substantial mass transfer from the smaller to the larger object 
\citep[see also Fig.~\ref{fig:lab}]{2010ApJ...725.1242K,2011Icar..215..596T,2009A&A...505..351T,2009MNRAS.393.1584T,2005Icar..178..253W}. 
Recent numerical simulations indicate that the gap between the onset of 
bouncing and the onset of the mass-transfer process can be bridged 
\citep{2012A&A...540A..73W,2012A&A...544L..16W,2013ApJ...764..146G} 
so that growth to planetesimal sizes seems to be possible.
As most of the laboratory experiments have been performed with 
silicates, the above statements might not be applicable to icy 
particles, i.e. for distances to the central star beyond the snow line. 
\citet{2012ApJ...752..106O} 
studied the mass evolution of aggregates consisting of 0.1~$\mu$m ice 
aggregates and found that there is no bouncing barrier so that icy 
planetesimals with extremely high porosity
(density below a few $10^{-4}$ g cm$^{-3}$) may be formed by direct collisional 
growth. 

\subsection{\bf Is planetesimal formation by direct sticking collisions possible?}

As indicated above, mass transfer from projectile to target agglomerate 
has been identified in the laboratory and confirmed in numerical 
simulations as a potential process by which in principle arbitrarily 
large dust aggregates, i.e. planetesimals, can be formed. This 
mass-transfer process in collisions at velocities at which the smaller 
of the dust aggregates fragments, can continually add mass to the 
growing larger dust aggregate as long as the absolute size and impact 
velocity of the (smaller) projectile aggregate is below a threshold 
curve derived by \citet{2009MNRAS.393.1584T}. However, 
\citet{2011ApJ...734..108S} showed in laboratory experiments that there 
is also a lower threshold mass for the mass-transfer process to be 
active (see Fig.~\ref{fig:lab}). When growing dust aggregates are 
continuously bombarded by monomer dust grains of micrometer size, 
they can be eroded quite efficiently. Thus, the formation of 
planetesimals by mass transfer can only be possible if the mass 
distribution of dust aggregates does not favor erosion (too many 
monomer dust grains) or fragmentation (too many similar-sized large 
dust aggregates). Much more work, both in the laboratory and with 
numerical simulations, is required before the question whether 
planetesimals can form by collisional sticking can be finally answered.

%
%
\section{\textbf{MODELS OF DUST EVOLUTION IN DISKS}}
\label{Sec:Theo}

\subsection{Introduction and early works}
The previous sections have laid the basis of dust evolution, such as transport processes, collision
velocities, and collisional outcomes. In this section, we will now bring all of this together and
review how grains in protoplanetary disks grow, how they are transported, and how growth and
transport effects work with or against each other.

\medskip

Some of the earliest works on dust particle growth in the context of planet formation were done by
\citet{Safronov:1969p11177}, who considered both a toy model (growth of a single particle sweeping
up other non-growing particles) as well as analytical solutions to the equation, which governs the
time evolution of a particle size distribution, often called the Smoluchowski equation
\citep{Smoluchowski:1916p2203}. A rather general way of writing this equation is
\begin{small}
\begin{equation}
\dot n(x)  = \iint_{x_1,x_2}^\infty n(x_1) n(x_2) K(x_1,x_2)
L(x_1,x_2,x)\mathrm{d}x_1\mathrm{d}x_2.
\label{eq:theo:smoluchowski}
\end{equation}
\end{small}
Here $x$ denotes a vector of properties such as composition, porosity, charge, or others, but to
keep it simple, we can think of it just as particle mass $m$. Eq.~\ref{eq:theo:smoluchowski} means
that the particle number density $n(m)$ changes if particles with masses $m_1$ and $m_2$ collide at
a rate of $K(m_1,m_2)$, where $L(m_1,m_2,m)$ denotes the mass distribution of the collisional
outcome. For example, for perfect sticking, $L(m_1,m_2,m)$ is positive for all combinations of $m_1$
and $m_2=m-m_1$, creating a particle of mass $m$, negative for all cases where $m=m_1$ or $m=m_2$,
and 0 for all other combinations. Care has to be taken that the double integral does not count each
collision twice. In principle, the collision rate $K$ and outcome $L$ can be a function of other
properties as well, but in most applications the collision rate is simply a product of collisional
cross section and relative velocity.

\medskip

The early works of \citet{Safronov:1969p11177}, \citet{1980Icar...44..172W}, and
\citet{Nakagawa:1981p4533} considered dust grains settling towards the mid-plane and sweeping up
other dust grains on the way. It was found that within few $10^3$ orbits, cm-sized dust grains can
form near the mid-plane and the gravitational instability of this dust layer was investigated.
\citet{1984Icar...60..553W} showed that turbulence could cause high collision velocities and lead to
break-up of aggregates instead of continuing growth.

\medskip

Numerical modeling of the full coagulation equation was found to be very difficult, as the dynamical
range already from sub-micrometer to decimeter sizes spans 18 orders of magnitude in mass and it was
shown that a rather high number of mass sampling points is needed, in order to reach agreement
between numerical and analytical results \citep[e.g.][]{Ohtsuki:1990p799,Lee:2000p11145}. Many works
therefore considered simplified models that use averaged values and only a single size.

\medskip

In the following, we will focus on more recent works, first, local models of collisional dust
evolution, followed by simplified and full global models. For more historical reviews on the
subject, we refer to previous reviews of \citet{1993prpl.conf.1031W}, \citet{Beckwith:2000p20930},
or \citet{Dominik:2007p1420}.

\subsection{Recent growth models and growth barriers}\label{theo:modeling:zero_dimensional}
Due to the complications mentioned above, modeling dust growth, even only at a single point in space
can be challenging, depending on which effects and parameters are taken into account.
One of the most detailed methods of dust modeling is the N-body like evolution of monomers, as done
by \citet{1999Icar..141..388K}, however it is impossible to use this method for following the growth
processes even only to mm sized particles, due to the sheer number of involved monomers.

\medskip

Monte-Carlo methods make it computationally much more feasible to include several additional dust
properties, such as porosity, and were first applied in the context of protoplanetary disks by
\citet{2007A&A...461..215O,2008ApJ...679.1588O}. Similar Monte-Carlo methods as well as the
experiment-based, detailed collision model of \citet{2010A&A...513A..56G} were used in
\citet{2010A&A...513A..57Z}, which followed the size and porosity evolution of a swarm of particles.
In agreement with previous studies, the initial growth occurs in the hit-and-stick phase which leads
to the formation of highly porous particles. As particle sizes and impact velocities increase, the
main collision outcome shifts to bouncing with compaction, in which two colliding particles do not
stick to each other, but are only compacted due to the collision. The final outcome in these models
was a deadlock where all particles are of similar sizes and the resulting impact velocities are such
that only bouncing collisions occurred, i.e., further growth was found to be impossible and this
situation was called the \textit{bouncing barrier}.

\medskip

\begin{figure*}[thb]
\begin{center}
  \includegraphics[width=8.5cm]{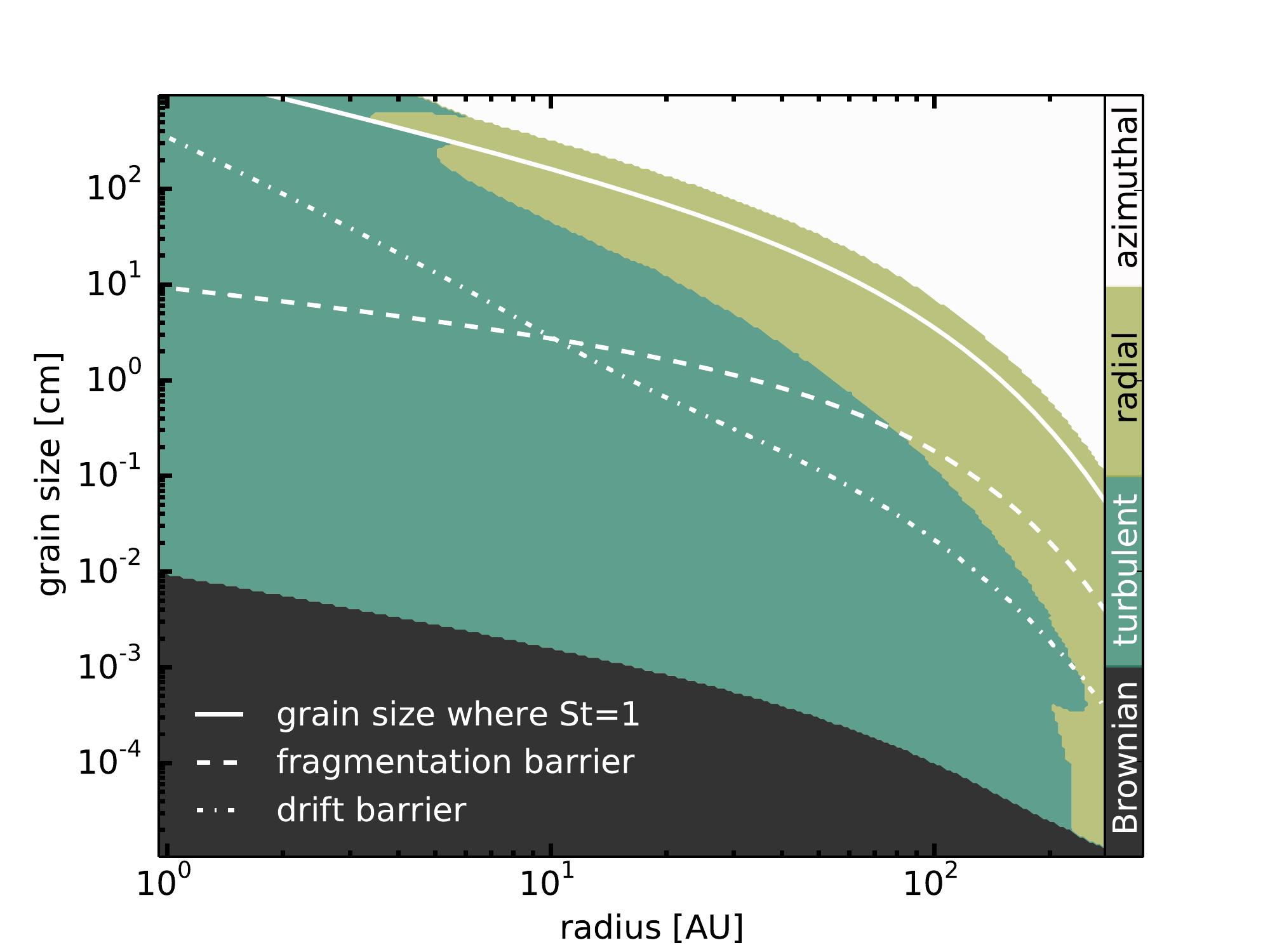}
  \includegraphics[width=8.5cm]{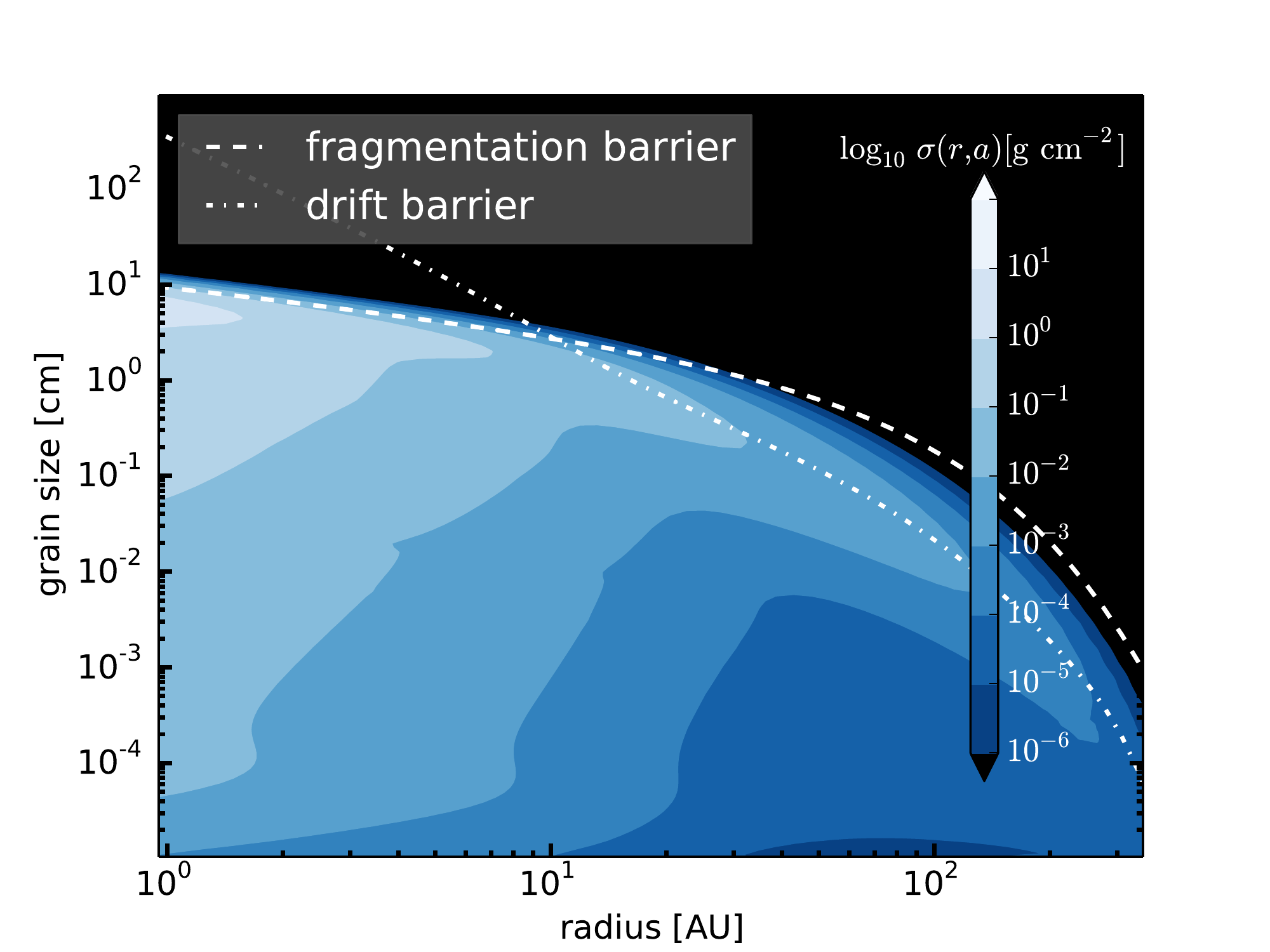}
  \caption{Left panel: important particle sizes and regimes of relative velocities in a fiducial disk
  model. The solid white line denotes the particle size corresponding to a Stokes number of unity,
  i.e. the fastest drifting dust particles. The dashed and dash-dotted lines are the size barriers
  due to fragmentation and radial drift, respectively (see Eqs.~\ref{eq:theo:a_frag} and
  \ref{eq:theo:a_drift}). The colored areas denote the dominating source of relative velocities at a
  given particle size and radius. This plot used a disk mass of 1\% of the stellar mass, a stellar
  mass of 1~$M_\odot$, a turbulence parameter $\alpha_\mathrm{t}$ of $1\times 10^{-3}$, a
  fragmentation velocity of 3 m~s$^{-1}$, a solid density of 1.6 g~cm$^{-3}$, and a self similar
  gas surface density \citep[see][]{LyndenBell:1974p1945} with viscosity index p=1 and
  characteristic radius of 60~AU, and a temperature profile
  of $T=50 K (r/10~\mathrm{AU})^{-0.5}$. Right panel: grain size distribution as a function of
  radius on the disk after $3 \times 10^6$~years for the same parameters as above. The initial
  dust-to-gas ratio is $10^{-2}$, but the resulting dust-to-gas ratio at 3 Myrs was used in the left
  panel.}
  \label{fig:theo1}
\end{center}
\end{figure*}

Even if bouncing is not considered, and particles are assumed to transition from sticking directly
to a fragmenting/eroding collision outcome, (as in the statistical Smoluchowski models of
\citealp{2008A&A...480..859B}), a similar problem was found: as particles become larger, they tend
to collide at higher impact velocities (see Sect.~\ref{theo:growth:relative_velocities}). At some
point the impact velocity exceeds the threshold velocity for shattering. Equating this threshold
velocity above which particles tend to experience disruptive collisions with an approximation of the
size-dependent relative velocity, yields the according size threshold of
\citep[see,][and Fig.~\ref{fig:theo1}]{2012A&A...539A.148B}
\begin{equation}
a_\mathrm{frag} \simeq \frac{2}{3\pi}\frac{\Sigma_\mathrm{gas}}{\rho_\mathrm{s}\,\alpha_\mathrm{t}}
\frac{u_\mathrm{frag}^2}{c_\mathrm{s}^2},
\label{eq:theo:a_frag}
\end{equation}
where $\Sigma_\mathrm{gas}$ is the gas surface density, $\rho_\mathrm{s}$ the material density of
the dust grains and $u_\mathrm{frag}$ the fragmentation threshold velocity derived from lab
measurements. Earlier models were using energy arguments to decide whether fragmentation happens
\citep[e.g.][]{Weidenschilling:1997p4593}, or did not include the effects of erosion
\citep[e.g.][]{2005A&A...434..971D}. More recent models of \citet{2012A&A...544L..16W}, which use
an experiment-based collisional outcome model indicate that most of the dust mass is still expected to remain in
small, observationally accessible particles ($\lesssim$1~cm) while allowing the formation of a small
number of large bodies.

\medskip

If fragmentation is limiting further growth of particles, a steady-state distribution is quickly
established in which grains undergo numerous cycles of growth and subsequent shattering, i.e. growth
and fragmentation balance each other. The resulting grain size distributions tend to be power-laws
or broken power-laws, which do not necessarily follow the MRN-distribution
\citep[see,][]{2011A&A...525A..11B}. A common feature of these distributions is the fact that most
of the dust mass is concentrated in the largest grains. However, the total dust surface area is
typically dominated by small grains of about a few micrometers in radius. Below a few micrometers,
the main source of collision velocities switches from turbulence to Brownian motion
(see Fig.~\ref{fig:theo1}), which leads to a kink in the size distribution, such that particles much
smaller than a micrometer do not contribute much to the total grain surface area
\citep[see,][]{2011A&A...525A..11B,Ormel:2013p21040}.

\medskip

Another recent development in dust modeling was the introduction of a method to self-consistently
evolve the porosity in addition to the size distribution with the conventional Smoluchowski method,
by \citet{Okuzumi:2009p9772}. Including another particle property, such as porosity, increases the
dimensionality of the problem from one dimension (mass) to two dimensions (mass and porosity), which
makes conventional grid-based methods prohibitively slow. By including an additional property in
the Smoluchowski equation and assuming it to be narrowly distributed, \citet{Okuzumi:2009p9772}
derived the moment equations, which describe the time evolution of the size distribution
and the time evolution of the mean porosity of each particle size. These moment equations are
one-dimensional equations of the particle mass, which can be solved individually with the
conventional Smoluchowski solvers. The results of this method showed good agreement with Monte-Carlo
methods at significantly increased computational speeds.

\medskip

This method was also used for including grain charge in \citet{Okuzumi:2011p15377} to confirm a
scenario of \citet{Okuzumi:2009p7473}, in which dust grain charging and the subsequent repelling
force could halt grain growth. This was termed \textit{charging barrier}. They found general
agreement with their previous scenario. In particular, they found that considering a distribution of
relative velocities (for Brownian motion and turbulence), in contrast to assuming that all grains
collide with the r.m.s. velocity, does not allow growth beyond the charging barrier. If turbulent
mixing is considered, however, the charging barrier could be overcome \citep{2011ApJ...731...96O}.

\medskip

The effects of velocity distributions were taken into account in \citet{2012A&A...544L..16W} and
\citet{2013ApJ...764..146G} to show that also the bouncing barrier can be overcome: not all
particles collide with the r.m.s. velocity. Instead, also collisions at higher velocities (possibly
causing fragmentation or erosion) or lower velocities (possibly causing sticking) occur. These
additional outcomes open up a channel of growth as some particles might always be in the ``lucky''
regime of velocity space and continue to stick, while all particles can also sweep up the fragments
produced by high-velocity impacts. In this sense, when a distribution of collision speeds and/or
turbulent mixing is taken into account, both the charging as well as the bouncing barrier become
``porous barriers''. They can slow down particle growth, but they cannot prevent the formation of
larger bodies.

\medskip

Even if particles beyond the usual growth barriers are formed, i.e., particles with sizes of around
centimeters or meters, depending on the position in the disk, it still seems unrealistic to
assume that these boulders stick to each other at \textit{any} velocity. Laboratory results of
\citet{2005Icar..178..253W}, however might give a solution to this problem: it was found that
impacts of small grains onto larger targets at velocities of the order of 25-50 m~s$^{-1}$ still
lead to net growth of the target, while some of the projectile mass is redistributed to smaller
fragments \citep[see also][]{2009A&A...505..351T,2009MNRAS.393.1584T,2010ApJ...725.1242K}. This
mechanism of \textit{fragmentation with mass transfer} was shown to be a way in which larger
particles, embedded in a large number of small dust grains, can continue to grow by sweeping up the
small grains at high impact velocities \citep{2012A&A...540A..73W}.

\subsection{Simplified global models}

Modeling the evolution of a full particle size distribution faces many conceptual challenges and
comes at significant computational costs. These complications have led to several models of the
global evolution of dust in protoplanetary disks that focus on the transport side of the evolution
and use only a simplified size evolution.

\medskip

Some earlier works focused only on the very early evolution or on a limited radial range
\citep[e.g.][]{1984ApJ...287..371M,1994Icar..111..536S}. Models of
\citet{1996A&A...309..301S,1997A&A...319.1007S}, and \citet{Hughes:2012p20952} followed also the
mono-disperse growth of dust particles, however they only considered pure sticking, i.e. particles
could grow without limits. They showed that particles can be quickly depleted from the disk if they
grow and drift. While in \citet{2002ApJ...580..494Y}, particles of a fixed size accumulate because
the drift velocity becomes slower in the inner regions of the disk, in the case of
\citet{1997A&A...319.1007S} and also \citet{Kornet:2001p688}, particles grow as they drift inward,
which counteracts the decrease in velocity. In the case of a massive disk, the entire dust
population is lost towards the inner boundary. Similar results were found by
\cite{2008A&A...487..265L}, who implemented the growth prescription of \citet{1997A&A...319.1007S}
into an SPH code.

\medskip

\citet{2007ApJ...671.2091G} extended upon this idea by assuming the particle size distribution to be
a MRN-like power-law \citep[see,][]{Mathis:1977p789} up to a certain maximum size. The maximum size
was assumed to evolve as it sweeps up other grains but the small grains were not evolved
accordingly, i.e. they were assumed to be produced by fragmentation, even though fragmentation was
not taken into account. This way, similar to previous works, particles could grow unhindered to
planetesimals. One important finding of \citet{2007ApJ...671.2091G} was that the dust mass in the
outer disk represents a reservoir for the inner disk. As the small grains in the outer disk grow,
they start to spiral inwards due to radial drift and feed the inner disk.

\medskip

Another approach was taken by \citet{2012A&A...539A.148B}, where results of full numerical
simulations (see Sect.~\ref{theo:modeling:full_models}) were taken as a template for a simplified
model. In this model, the dust distribution is divided into small grains, which are basically
following the gas motion, and large grains which contain most of the mass and are subject to
significant radial drift. The full numerical results showed that the shape and upper limit of
the size distributions mostly represent two limiting cases, where either radial drift or grain
fragmentation is the size limiting factor. The resulting grain size yields an effective transport
velocity that can be used to calculate the evolution of the surface density, which agrees well with
the full numerical solutions.

\subsection{Detailed global dust evolution models}\label{theo:modeling:full_models}
The first models treating both the dust collisional evolution and radial transport in a
protoplanetary disk were done by \citet{1997A&A...325..569S} and \citet{2001ApJ...551..461S},
however these works could only simulate the evolution for $100$ and $10^4$ years, respectively.
Still, they already showed that especially in the innermost regions, grain growth is occurs very
quickly and that the initial growth is driven by Brownian motion until turbulence induced relative
velocities dominate at larger sizes.

\medskip

The models of \citet{2005A&A...434..971D} and \citet{2005ApJ...625..414T} were global in the sense
that they simulated dust evolution and vertical transport at several radial positions of the disk,
thus being able to calculate spectral energy distributions of disk, but they did not include the
radial transport.

\medskip

The models of  \citet{2008A&A...480..859B} and \citet{2010A&A...513A..79B} were among the first to
simulate dust evolution and radial transport for several millions of years, i.e. covering the
typical life time of protoplanetary disks. This was possible by treating the dust evolution in a
vertically integrated way, which reduces the problem to one spatial and the mass dimension. They
showed that both radial drift and grain shattering collisions pose a serious obstacle towards the
collisional formation of planetesimals. \citet{2008A&A...487L...1B} used such a model to show that
in the scenario of \citet{Kretke:2007p697}, dust particles can accumulate in pressure bumps and form
larger bodies, as long as turbulence is weak enough to avoid particle fragmentation.

\medskip

A common feature of basically all simplified and full global models is the \textit{radial drift
barrier}, which limits the maximum size that particles can reach. Technically speaking, it is not a
growth barrier, as it is not limiting the particle growth itself, but it still enforces an upper
size cut-off by quickly removing particles larger than a particular size. Setting aside the other
possible barriers and assuming particles perfectly sticking upon collision, we can define a growth
time scale $t_\mathrm{grow} = a/\dot a$, which is the time scale on which the size is doubled. For a
mono-disperse size distribution of spherical grains, the growth rate can be written as $\dot a =
\Delta u\, \rho_\mathrm{dust}/\rho_\mathrm{s}$, where $\rho_\mathrm{dust}$ is the dust density and
$\Delta u$ the collision velocity. $\dot a$ can be thought of as the velocity of motion along the
vertical axis in Fig.~\ref{fig:theo1}, while the radial drift velocity is the motion along the
horizontal axis. The size limit enforced by drift is therefore approximately where the growth time
scale exceeds the drift time scale $t_\mathrm{drift} = r/u_\mathrm{drift}$, which is approximately
given by \citep[see,][and Fig. ~\ref{fig:theo1}]{2012A&A...539A.148B}
\begin{equation}
a_\mathrm{drift} \simeq 0.35 \frac{\Sigma_\mathrm{dust}}{\rho_\mathrm{s}}
\frac{V_\mathrm{K}^2}{c_\mathrm{s}^2} \left| \frac{\mathrm{d ln}P}{\mathrm{d ln}r}\right|^{-1}.
\label{eq:theo:a_drift}
\end{equation}

\medskip

Having most of the dust mass contained in the largest grains of a given size $a_\mathrm{max}(r)$,
the transport velocity of the dust surface density can roughly be approximated as
$u_\mathrm{drift}(a_\mathrm{max}(r))$. In a quasi-stationary situation, where the dust is flowing
inward at a rate $\dot M_\mathrm{dust}$, the surface density profile follows directly from the mass
conservation
\begin{equation}
\Sigma_\mathrm{dust} = \frac{\dot M_\mathrm{dust}}{2\pi\,r\,u_\mathrm{drift}}.
\label{eq:theo:sig_dust}
\end{equation}
If the largest grain size is set by fragmentation, which is mostly the case in the inner disk (cf.
Fig.~\ref{fig:theo1}), then the dust surface density was shown to follow a profile of
$\Sigma_\mathrm{dust} \propto r^{-1.5}$, which is in agreement with both the estimates for the solar
system \citep{Weidenschilling:1977p15694,Hayashi:1981p15696} and the estimates from extrasolar
planets of \citet{Chiang:2013p20955}.

\medskip

If radial drift is at play, and the dust surface density drops, then particularly the outer disk
will at some point become dominated by the radial drift barrier, in which case the resulting dust
surface density profile was shown to be $\Sigma_\mathrm{dust} \propto r^{- (2p+1)/4}$ (where
$\Sigma_\mathrm{gas} \propto r^{-p}$), in agreement with current observations of
\citet{andrews:2012}.

\medskip

A possible way of overcoming the drift barrier, however for a limited spatial range, was discussed
in \citet{2012ApJ...752..106O}. They used the method of \cite{Okuzumi:2009p9772} to simulate the
porosity evolution in addition to the global evolution of non-fragmenting dust particles, based on a
collisional model of \citet{Suyama:2012p20954}. They found that outside the snow line, but inside of
around 10~AU, extremely porous particles, as formed in this collisional model (internal densities of
the order of $10^{-5}$ g~cm$^{-3}$), have a strongly decreased growth time due to their increased
collisional cross section, which allows them cross the drift barrier. Sintering
\citep[e.g.][]{Sirono:2011p21038,Sirono:2011p21039} is expected to prevent this mechanism inside
approximately 7~AU \citep{2012ApJ...752..106O}. However, the precise impact that sintering, or
fragmentation has on this mechanism remains to be investigated.

\subsection{Summary: Modeling}
Current modeling efforts are standing on the proverbial shoulders of giants: 
laboratory work, numerical modeling of individual particle collisions, 
and theoretical studies on disk structure, turbulence, collision velocities, 
and other physical effects represent the foundation on which global models 
of dust evolution are build. 

\medskip

Models predict grains to grow to different sizes, migrate and mix radially
and vertically throughout the disk. Large grains are expected to settle
on the mid-plane and be transported radially to the inner disk regions, 
while small grains are mixed vertically to the disk surface by turbulence and 
transported outwards as the gas spreads out as part of the disk viscous evolution.
The growth timescale is also longer in the outer regions of the 
disk, as a consequence, this region acts as a reservoir for the inner disk.
Dust evolution in the planet forming regions of the disk is strongly
influenced by the rate at which large solids are transported inwards from the 
outer disk. Understanding how to slow down the radial transport is one of 
the important goals for the future research in this field.

\medskip

Current models of non-fractal grain growth can explain the formation 
of larger bodies by incremental growth, but most of the dust mass tends 
to be trapped either by fragmentation or by radial drift. Size limits 
associated with these effects lead to a
characteristic dust surface density distribution, that can be compared 
directly with observations (see \S~\ref{Sec:Obs}).
Fractal dust aggregates with internal densities of the order of 
$10^{-5}$ -- $10^{-3}$ g cm$^{-3}$ experience accelerated growth and could 
break through the radial drift barrier. A detailed analysis
of the evolution of these aggregates and their observational 
properties are an active area of research, where we also expect
rapid progress in the coming years.

%
%
\section{\textbf{EFFECTS OF GRAIN GROWTH ON DUST OPACITY}}
\label{Sec:Opacities}

The wavelength-dependent absorption and scattering properties of dust grains 
change as they grow in size. This behavior provides a tool to study the grain 
growth through panchromatic observations of circumstellar disks from optical 
to centimeter wavelengths. 

A significant fraction of the disk emission at optical and near-infrared 
wavelengths consists of stellar light scattered by the dust grains in the 
superficial layer of the disk (see Fig~\ref{fig:sketch}).  
The morphology and  color  of the scattered light is most sensitive 
to grains with sizes between 0.01-10~$\mu$m, which correspond to the 
transition between the Rayleigh and the Mie scattering. As a general trend, 
as grains grow the scattering phase function becomes more isotropic and 
the color of the scattered light turns redder \citep{1983asls.book.....B}. 
In addition, the strength 
of mineral spectroscopical features, e.g. the silicate resonance feature at 10~$\mu$m, 
decreases \citep[see ][ and references therein]{2006ApJ...639..275K}.
As discussed in \S~\ref{Sec:scattir}, spatially resolved and spectroscopic observations of disks 
allow therefore to measure the size of dust grains in the surface layer of the disk.  

Observations at longer wavelengths probe the dust properties in the 
disk interior where most of the mass is located. From far-infrared to millimeter 
wavelengths, the disk emission is dominated by the thermal emission from 
warm dust which is controlled by the dust opacity $\kappa_{\nu}$.
For dust grains with minimum and maximum sizes
$a_{min}$ and $a_{max}$, a good approximation is a power law,
$\kappa_{\nu} \propto \nu^{\beta}$,
where the spectral index $\beta$ is sensitive to $a_{max}$ but does not depend on $a_{min}$ 
for $a_{min} < 1 \mu$m \citep{1993Icar..106...20M,2006ApJ...636.1114D}.
 
\begin{figure}[th]
\includegraphics[width=8.7cm]{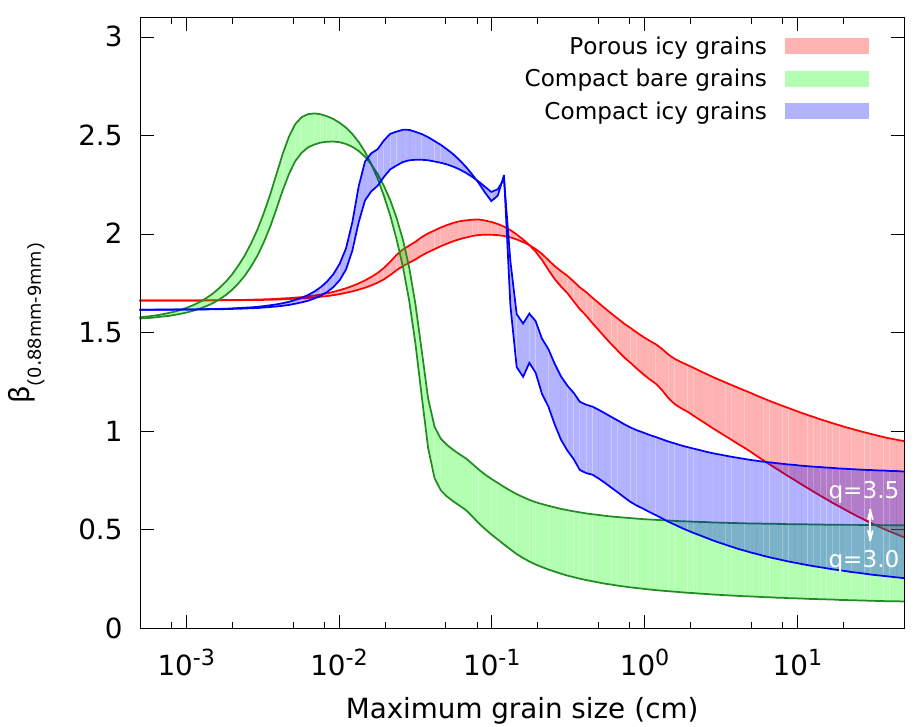}
\caption{Spectral index $\beta$ of the dust opacity $\kappa_\nu \propto \nu^{\beta}$ 
calculated between the wavelengths of 0.88~mm and 9~mm as a function of the maximum 
grain size, for a grain size distribution $n(a) \propto a^{-q}$ characterized by a minimum 
grain size of 0.01~$\mu$m. Different colors corresponds to grains with different chemical composition 
and porosity. Red: grains composed by astronomical silicates, carbonaceous material, and water ice, 
with relative abundances as in \citet{1994ApJ...421..615P} and a porosity of 50\%. Blue: compact grains with the 
same composition as above. Green: compact grains composed only of astronomical silicates and carbonaceous 
material. For each composition, the colored region shows the values of $\beta$ in the range $q=3.0$ to $q=3.5$.
Despite the dependence of $\beta$ on the grain composition and the value of $q$, maximum grain sizes larger 
than about 1~mm lead to values of $\beta$ less than unity. This opacities have computed following the prescription as in \citet{2004ASPC..323..279N}.
}
\label{fig:beta}
\end{figure}

The dust opacity spectral index $\beta$ does not depend only on the 
grain sizes of the emitting dust, but also on other factors such as 
dust chemical composition, porosity, geometry, as well as on the
the grain size distribution, normally assumed as a power law of the form
$dN = n(a)^{-q} da$ \citep[e.g.][]{2007prpl.conf..767N}. 
The main result is that, regardless of all the uncertainties 
on the dust model, dust grains with sizes of the order of 1 mm or larger lead to 
$\beta$ values lower than about 1 
\citep[Fig.~\ref{fig:beta}, see also][]{2004ASPC..323..279N}.

The inferred range of $\beta$ values in young disks has also an important 
consequence on the derivation of dust masses through sub-mm/mm photometry, 
which implies  knowledge of the dust opacity coefficient. Solids with 
sizes much larger than the wavelength of the observations do not 
efficiently emit/absorb radiation at that wavelength and are therefore 
characterized by low values of the dust opacity coefficient. 
Physical models of dust emission have been used to quantify the effect 
of changing $\beta$ on the millimeter dust opacity coefficient. For example, 
for the dust models of \citet{2006ApJ...638..314D} with a slope of 2.5 
for the grain size distribution, a dust population with $\beta \approx 0.2$ 
has a dust opacity coefficient at 1~mm lower by nearly 2 orders of magnitude 
than a dust population with $\beta \approx 1.0 - 1.5$. This difference can be understood 
by the fact that, for a given slope of the grain size distribution, a  value of
$\beta$ much lower than 1 requires  extending the distribution to very large grains, 
much larger than the observing wavelength. Similar results are 
obtained with the dust models considered in \citet{2010A&A...512A..15R}.  
Despite the fact that the absolute value of the dust opacity at any wavelength strongly 
depends on the adopted dust model, this shows how inferring relations which 
involve disk dust masses without accounting for possible variations of the 
$\beta$ parameter throughout the sample can lead to potentially large biases 
and errors.

%
%
\section{\textbf{DUST EVOLUTION BEFORE THE DISK FORMATION}}
\label{Sec:pre-disc}

\noindent
This review focuses on the dust evolution in protoplanetary discs as the
first step of the formation of planetesimals and larger bodies in 
planetary systems. Nevertheless, many of the physical processes described
in the first two sections are also expected to occur in molecular cloud cores 
and in the envelopes of protostars before the disk formation stage. 

\medskip

Grain coagulation and growth in cores and protostellar envelopes have been 
modeled by several authors 
\citep[e.g.][]{1993A&A...280..617O,1994ApJ...430..713W,1999ApJ...524..857S,2009A&A...502..845O}.
These models show that grains can form fluffy aggregates very efficiently as 
a function of core density and 
time. The growth is favored by the presence of ice mantles, which are 
expected to form in the denser and cooler interior of the clouds. 
Taking the most recent calculations by \citet{2009A&A...502..845O}, 
on a timescale of $\sim$1~Ma, grains can grow to several micron-size 
particles at densities of $\sim$10$^5$\,cm$^{-3}$, and reach several 
hundred micron size aggregates at higher densities. The effects of 
these changes on the dust scattering and absorption properties have 
also been computed and has been shown that the growth can be traced combining
absorption, scattering and emission at different wavelengths 
\citep{1994A&A...288..929K,1994A&A...291..943O,2011A&A...532A..43O}.

\medskip

Detailed modeling of the infrared scattered light from dark cores in  
molecular clouds has shown widespread evidence for the presence of
dust grains grown to several micron size particles, 
even in relatively low density regions \citep{2010Sci...329.1622P,2010A&A...511A...9S}.
This is consistent with the models described above if the cores survive 
for timescales of the order of $\ge$1~Ma. These findings are in 
agreement with studies of the optical and infrared extinction law
in nearby molecular clouds \citep[eg.][]{2013MNRAS.428.1606F}.
Evidence of growth to larger sizes in the densest regions of molecular
cloud cores have been investigated by measuring the dust emission at far 
infrared and submillimeter wavelengths and comparing the emission properties with 
extinction in the optical and infrared 
\citep[eg.][]{2003A&A...398..551S,2013ApJ...763...55R,2013A&A...555A.140S}. 
Solid results in this context have proven elusive so far, the main 
limitations of the current studies are the sensitivity of wide area infrared 
surveys that do not 
see through the densest regions of cores. The use of the sub-millimeter spectral
index as a probe of the grain size distribution has been questioned 
by several studies which showed a possible dependence of the dust
opacity with temperature \citep{2010A&A...520L...8P,2010ApJ...713..959V}. 
However, recent detailed studies of dense and 
cold clouds which include a broader range of (sub-)millimeter wavelengths, suggest
that, in some cases, the observed anti-correlation between the dust opacity spectral
index $\beta$ and temperature may be the result of the uncertainties resulting
from using simplified single temperature modified black body fits to observations
covering a limited range of wavelengths 
\citep{2009ApJ...696.2234S,2012ApJ...752...55K,2013ApJ...767..126S,2013A&A...556A..63J}.
Indirect evidence supporting grain growth in the dense regions of pre-stellar 
cores has been obtained by \citet{2008ApJ...683..238K}, which invoke significant
grain growth at densities exceeding 10$^5$~cm$^{-3}$ to explain the inferred change
in dust opacity in the inner regions of the studied pre-stellar cores.

\medskip

A recent development has been the study of dust properties in protostellar 
envelopes and young disks. The dust opacity index in the youngest protostars 
indicate possible evidence for large grains  in the collapsing envelopes
\citep{2009A&A...507..861J,2009ApJ...696..841K}. These initial results
have been recently followed up with better data and more detailed modeling,
which confirmed the initial suggestion that $\beta\le 1$ in the inner envelopes of 
the youngest protostars \citep[][{\it Miotello, priv. comm.}]{2012ApJ...756..168C,2013ApJ...771...48T}.
These findings suggest that large dust aggregates, perhaps up to millimeter size, can form 
during the disk formation stage in the infalling envelopes and are broadly in agreement with a very fast formation
of Calcium-Aluminum Inclusions in carbonaceous chondritic meteorites 
\citep[][see also the chapter by Johansen et al.]{2012Sci...338..651C}.

%
%
\section{\textbf{CONSTRAINTS ON DUST GROWTH IN DISKS FROM NIR AND MIR OBSERVATIONS}}
\label{Sec:scattir}


Because of the impact of density on time scales for growth, it is
clear that the major part of dust growth has to happen in the mid-plane
of the disk, where densities are highest.  The outer surfaces of the
disk are mainly acting as display cases where products of this dust
growth, more or less modified by transport processes from the mid-plane
to the surface, are displayed to the observer by their interaction
with optical, near and mid IR radiation.  These surfaces are the upper
disk surface, at the inner rim.  The upper disk surface is relevant
because it is directly illuminated (if the disk is flaring), and it
also allows radiation at wavelengths well below the sub-mm regime to
escape.  As vertical mixing happens on time scales that are short
compared to radial transport, grains in the disk surface are related to
the population on the mid-plane at the same distance from the star.  
The inner rim of the disk is important because
dust gets exposed, often at temperatures close to its evaporation
temperature.  The inner rim also allows in principle to study mid-plane
dust by constraining the evaporation surface, the initial results show that
grains larger than micron-size particles are required, but this avenue 
of research has so far been only marginally explored 
\citep[e.g.][]{2006A&A...451..951I,2010ARA&A..48..205D,2013ApJ...775..114M}.

\medskip

The tracers of dust size at these surfaces are (i) Angular and
wavelength dependence of scattered light, both intensity and
polarization, and (ii) The shape and strength of dust emission
features \citep[][]{2010ARA&A..48..205D}.

\subsection{\bf Scattered light}

Scattered light images can resolve the disks of nearby young
stars down to distances of about 20AU \citep[e.g.][]{2007ApJ...665..512A} from
the star, in some cases even down to about 10AU \citep{2012A&A...538A..92Q}.
Scattered light carries information about grain size in the intensity
of the scattered light, and in particular in the angular dependence of
the intensity, i.e. the phase function.  The brightness of scattered
light is often low, indicating grains with low albedo \citep{2007ApJ...665..512A,2010PASJ...62..347F}.
The color of disks can be redder than that of the star
\citep{2007ApJ...665..512A,2003AJ....126..385C,2008ApJ...682..548W}, an
effect that can be explained by the presence of large grains with
wavelengths-dependent effective albedo caused by a narrowing of the
forward-scattering peak at shorter wavelengths \citep{2012A&A...539A...9M}.
Observed disks often differ in brightness between the front and back
side \citep[e.g.][]{2008ApJ...673L..67K}, and a brighter front side is often taken
as a sign for large grains whose scattering phase function is
dominated by forward scattering \citep{2011ApJ...738...23Q}.  However, 
\citet{2013A&A...549A.112M} warn that large grains, if present in the disk surface,
may have phase functions with a very narrow forward scattering peak of
only a few degrees that can never be observed except in edge-on disks.
The behavior of the phase function at intermediate angles (the angles
actually seen in an inclined disk) may depend on details of the grain
composition and structure.

\medskip

Model fitting of scattered light images taken at multiple wavelengths
indicate the presence of stratification, with smaller grains higher
up in the disk atmosphere and larger grains settled deeper
\citep[e.g.][]{2007A&A...469..963P,2008A&A...489..633P,2010ApJ...712..112D}, in accordance with
the predictions of models \citep[e.g.][]{2004A&A...421.1075D}.

\subsection{\bf Dust features}

The usefulness of mid-IR  features as tracers of dust growth is
limited because the spatial resolution currently available at these
wavelengths makes it hard to study the shape of features as a function
of distance from the star, so only an average profile is
observed. Interferometric observations allow us to extract the spectrum
emitted by the innermost few AU \citep{2004Natur.432..479V}, showing
that the inner regions are much more crystallized than the disk
integrated spectrum indicated.  Pioneering work on the integrated
spectrum with ISO on Herbig stars
\citep{2001A&A...375..950B,2005A&A...437..189V} has in recent years
been extended to larger samples of Herbig stars, and to T Tauri stars,
making use of the better sensitivity provided by Spitzer. The result
of these studies is that grains of sizes around 1 to a few micrometers
must form quickly as the resulting broadening and weakening of the 10
and 20~$\mu$m features are present in most disks, essentially independent
of age and other stellar parameters
\citep{2006ApJ...639..275K,2006ApJS..165..568F,2010ApJ...721..431J,2011ApJ...734...51O}.
\textit{Juhasz et al.} find that larger grains are more prominent in 
sources that show a deficit of far infrared emission, interpreted as an
indication of a flatter disk
structure. These findings are in apparent contraddiction with the 
results of \citet{SiciliaAguilar:2007p20899}, who find an apparent 
trend for smaller grains in the atmospheres of older disks in Tr37. 
The interpretation for these observations is that larger grains settle 
more rapidly onto the midplane, as compared to small grains, and hence, 
at later ages, are not accessible to infrared observations. 
As infrared observations probe a thin surface layer of
the disk at a radius that is a strong function of 
the central star parameters \citep[eg.][]{2007ApJ...659..680K,2007prpl.conf..767N},
it is very difficult, and possibly misleading to use spatially unresolved spectroscopy as a probe of global dust growth in disks.
Nevertheless, an important result shown by all these studies is
that the dust producing the 10 and
20 micron features contains a significant fraction of crystalline
material, indicating that this dust was heated to temperatures of
around 1000~K during processing in the disk. 
\citet{2009Natur.459..224A} and \citet{2012ApJ...744..118J} clearly 
detected the increased production and subsequent transport to the 
outer disk of crystalline dust during the 2008 outburst of the star EX~Lup.
These findings suggest that eruptive phenomena in young disk-star systems
play an important role in the processing and mixing of dust in protoplanetary
disks.

\subsection{\bf Constraints for mid plane processes}

It is now generally believed that dust in the inner parts of the disk
quickly develops into a steady-state situation with on-going
coagulation and reproduction of small grains by erosion and
fragmentation (see section \ref{theo:modeling:zero_dimensional}).  The
dust present in the disk surface therefore does not seem to be a good
tracer of the evolution of the mean and maximum particles size in the
mid plane of the disk, but rather reports on the presence of this
steady state. While an  suggestion of a possible correlation 
between mid-plane and surface dust processing was made by 
\citet{2009A&A...495..869L}, many previous and subsequent attempts based on larger 
surveys have so far shown a distinct lack of correlation
\citep{2007prpl.conf..767N,2010A&A...521A..66R,2010ApJ...721..431J},
probably caused by the fact that mid infrared observations are tracing
the atmosphere of the inner disk (0.1-10~AU depending on the stellar
properties), while mid-plane dust has so far been probed only in the
outer disk (beyond $\sim 20$~AU, see \S~\ref{Sec:Obs}).

%
%
\section{\textbf{OBSERVATIONS OF GRAIN GROWTH IN THE DISK MID-PLANE}}
\label{Sec:Obs}

The denser regions of the disk mid-plane can be investigated 
at (sub-)millimeter and centimeter wavelengths, where dust 
emission is more optically thin. Long wavelength observations
are thus the unique tool that allow us to probe grain evolution 
on the disk mid-plane, where most of the mass of the solids 
is confined and where planetesimal and planet formation is 
thought to occur. In this section we will focus on 
the observational constraints on grain growth on the disk mid-plane
from long wavelength observations, describing the methodology and limitations 
of this technique (\S~\ref{Sec:Method}) as well as the results of the 
relatively extensive photometric surveys (\S~\ref{Sec:sub-mm_phot})
and of the more limited, but very promising for the future, spatially resolved 
multi-wavelength observations (\S~\ref{Sec:Resolved}).

\subsection{\textbf{Methodology}}
\label{Sec:Method}


As described in \S~\ref{Sec:Opacities}, as grains grow to sizes comparable 
with the observing wavelength, the dust opacity changes significantly,
imprinting the signature of growth in the disk emission. In particular,
at millimeter and centimeter wavelengths, the spectral index of the emission
of optically thin dust at a given temperature can be directly related to 
the spectral dependence of the dust opacity coefficient, which in turn 
is related to the maximum grain size.
 
Values of $\beta$ for dust in proto-planetary disks can be derived by 
measuring the slope $\alpha_{\rm{mm}}$ of the sub-mm SED 
($F_{\nu} \propto \nu^{\alpha_{\rm{mm}}}$). Under the simple 
assumptions of optically thin dust emission in the Rayleigh-Jeans tail 
of the spectrum, then $\beta = \alpha_{\rm{mm}} - 2$. 
The first single-dish and interferometric observations which measured 
the sub-mm slope of the SED of young disks came in the late 80s 
and early 90s 
\citep{1989ApJ...340L..69W,1989ApJ...337L..41W,1990ApJ...357..606A,1991ApJ...381..250B}. 
These pioneering works already revealed a wide interval of 
$\beta$-values ranging from $\approx 0$ up to the values typical 
of the ISM ($\approx 1.7$). The presence of dust grains with sizes 
larger than about 0.1~mm was soon recognized as a possible explanation 
for the disks showing $\beta < 1$ \citep{1989ApJ...340L..69W,1991ApJ...381..250B}.
 
\medskip

While this simplified approach has been useful in the initial studies,
it is obvious that it has serious limitations and should not be
used now that quick and efficient programs to self consistently 
compute the dust emission from a protoplanetary disk with an arbitrary
dust opacity are fast and widely available. All the results that we present
in this chapter are derived using disk models that include an approximate,
but proper treatment of the density and temperature profile in the disk and
use opacities computed from physical dust models 
\citep[e.g.][and successive improvements]{1997ApJ...490..368C,2001ApJ...560..957D,2007prpl.conf..555D,2004ASPC..323..279N}.
These models include an approximate treatment of the radiation transfer 
in the regions where the disk becomes optically thick even at millimeter 
wavelengths under the assumption of a smooth disk structure. The effect of 
including optically thick regions due to local over-densities at large 
distances from the star has been investigated and shown to be unlikely
to play a major role in protoplanetary disks by \citet{2012A&A...540A...6R}.
Another serious source of uncertainty, if proper disk emission
models are not considered, comes
from the assumption that the whole disk emission is in the 
Rayleigh-Jeans regime, as the dust temperature in the outer disk mid-plane
easily reach values below 15-20~K. 
Neglecting the effects of optical depth
and of the low dust temperatures can result in a significant 
underestimate of $\beta$, leading to incorrect results on grain growth 
\citep[see e.g.][]{1989ApJ...340L..69W,2001ApJ...554.1087T,2003A&A...403..323T,2005ApJ...626L.109W}.

\medskip

Another aspect to be considered is whether other sources of 
emission might be contaminating the signal from the large grains
at the observing wavelengths. The major source of uncertainties
comes from the contamination of the emission at long wavelength from 
gas in the stellar chromosphere or in the wind/jets. The typical 
approach used to resolve this uncertainty is to combine the 
millimeter observations with long wavelength observations that probe the
gas emission \citep{2001ApJ...554.1087T,2003A&A...403..323T,2005ApJ...626L.109W,2006A&A...446..211R}.
All these studies show that for T Tauri stars, not affected by intense 
external photoionization, the contribution of the gas emission at wavelength
shorter than 3~mm is below the 30\%\ level. It is important to note that this
is an estimate and, especially for the fainter disks and the higher mass
central stars the contribution may be significantly higher and need to be 
accounted properly. In regions where there is a strong external ionizing 
radiation, the contribution of the gas to the emission may be dominant
at 3mm and still be significant at 1mm. An example of this condition
are the inner regions of the Trapezium cluster in Orion
\citep[e.g.][and references therein]{2005ApJ...634..495W,2008ApJ...683..304E,2010ApJ...725..430M}. 
Since non-thermal emission and thermal free-free from an 
ionized wind vary on various time-scales 
\citep[e.g.][and references therein]{2010A&A...521A..32S,2012MNRAS.425.3137U}, 
nearly simultaneous observations at (sub-)mm and cm wavelengths 
should be used to quantify the gas spectrum and subtract it from 
the measured fluxes to infer the emission from dust only.

\subsection{\textbf{Results from multi-wavelegth sub-mm photometry}}
\label{Sec:sub-mm_phot}


\citet{2007prpl.conf..767N} covered the main results obtained in this 
field until the first half of the last decade. The main conclusion at the
time was that, while evidence for large grains had been found in several
bright disks, no clear trend with the properties of the system or age was
evident. Since then, many photometric surveys have targeted young disks in
several nearby star forming regions (distances $<$ 500~pc from the Sun) at  
long wavelengths, with the goal of constraining dust evolution in less 
biased samples.

\medskip

The most extensive studies aimed at a derivation of the grain growth properties 
have been performed for the Taurus-Auriga 
\citep{2005ApJ...631.1134A,2006A&A...446..211R,2010A&A...512A..15R} and
Ophiuchus \citet{2007ApJ...671.1800A,2010A&A...521A..66R} star forming
regions. Less extensive studies have also been carried out in southern
star forming regions \citep{2009A&A...495..869L,2012MNRAS.425.3137U} and
the more distant Orion Nebula Cluster 
\citep{2010ApJ...725..430M,2011A&A...525A..81R}.
It is important to note that all the surveys of disks conducted 
to date with the aim of characterizing the spectral index $\beta$ of 
the dust absorption coefficient are far from being complete in any star 
forming region, and the level of completeness decreases with decreasing 
disk mass.
In many cases the studies included a detailed analysis of the possible
contribution of the gas emission at long wavelengths and enough resolution 
to confirm that the disks are mostly optically thin and, for the large 
part, followed the methodology described in \S~\ref{Sec:Method} to derive
the level of grain growth.

\medskip

\begin{figure*}[th]
\includegraphics[width=8.5cm]{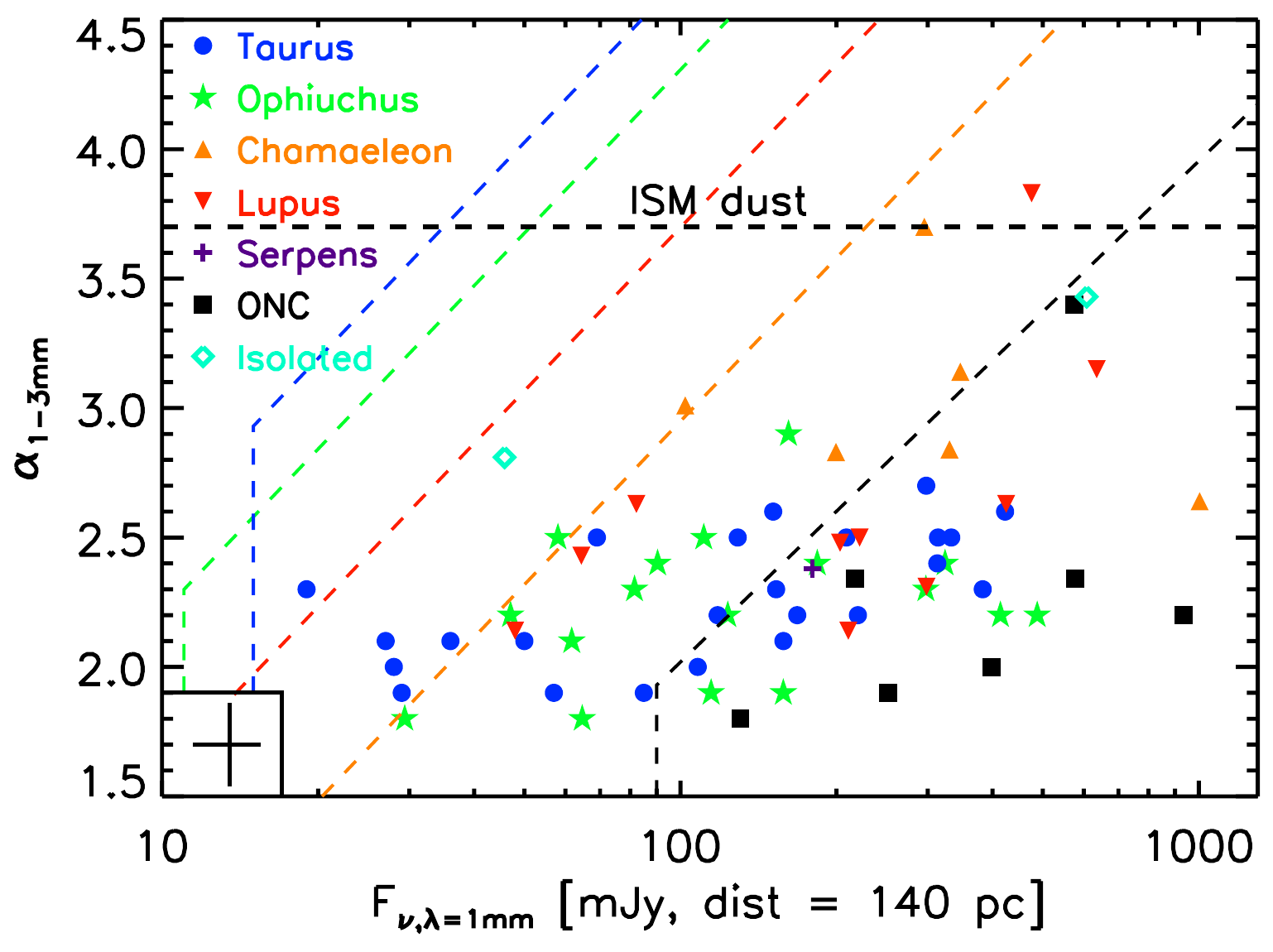}
\includegraphics[width=8.5cm]{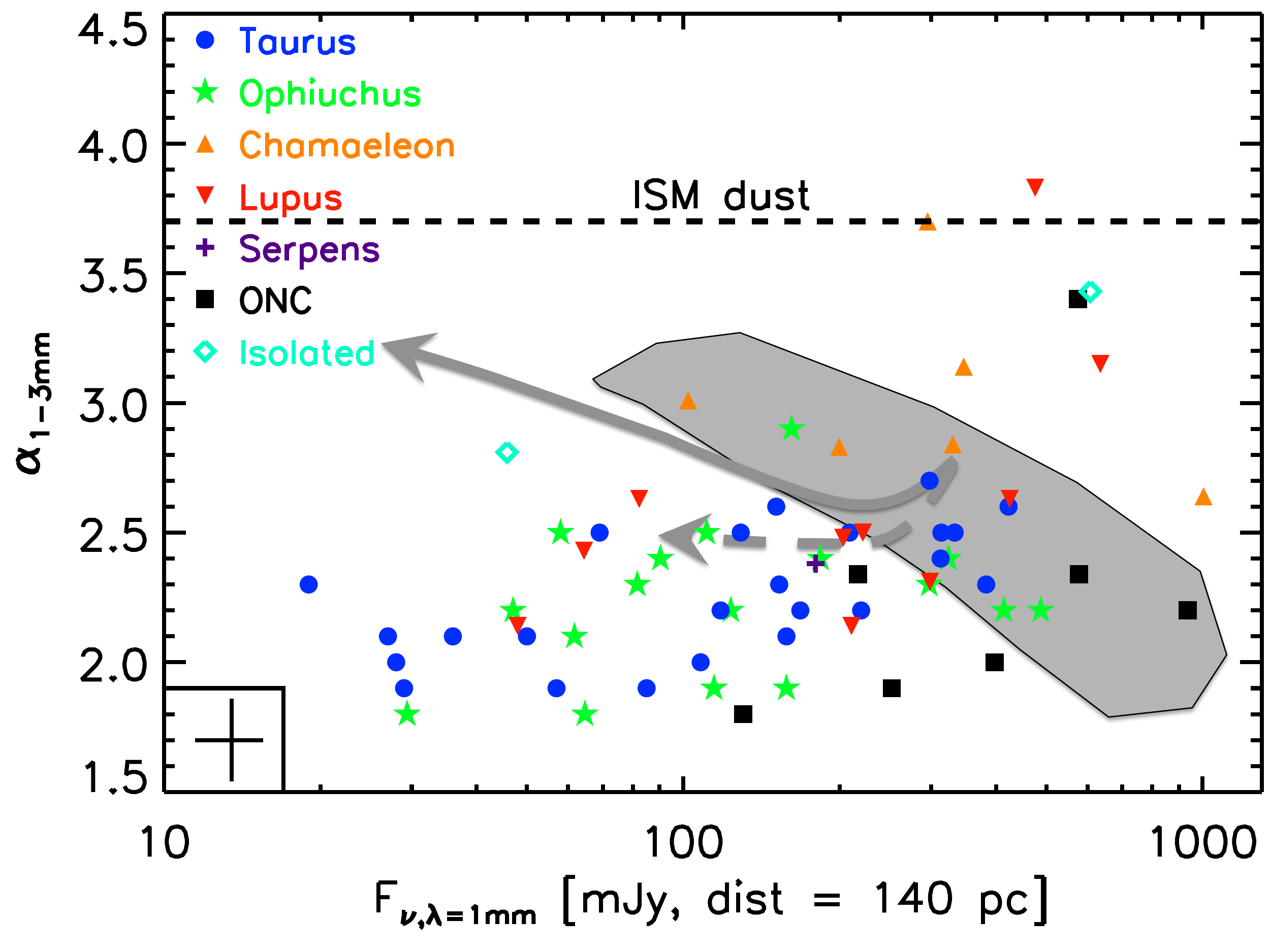}
\caption{Left panel: Spectral index between 1.1 and 3~mm plotted against the flux at 1.1~mm (scaled
for a common distance of 140~pc) for disks around single stars (or wide 
binaries) with spectral types early M to K in nearby star forming regions. The dashed 
lines mark the typical sensitivity limits of the surveys in Taurus, Ophiuchus, Lupus,
Chamaeleon and Orion Nebula Cluster.
Right panel: The grey area illustrate the range of predictions for global dust 
evolution models without radial drift \citep{birnstiel:2010}, the two arrows
illustrate the evolutionary trajectories in the first few million years as 
predicted by the global models including the effect of radial drift (solid line)
and including pressure traps in the gas distribution to slow the rate of drift
\citep[dashed line,][]{2012A&A...538A.114P}.
}
\label{fig:photometry}
\end{figure*}

Following the method laid out by \citet{2010A&A...512A..15R}, we
have selected a subsample of all the published measurements of the 
1.1--3~mm spectral index for disks surrounding single stars (or wide separation
binaries) of spectral type from K and early M. The measured spectral indices
are plotted in Fig.~\ref{fig:photometry} against the flux measured at 1.1~mm
(scaled at a common distance of 140~pc). The flux is roughly proportional to the 
total dust mass in the disk (assuming similar dust properties in all disks).

\medskip

The main results of these surveys is that the dust in the outer disk 
regions appears to have grown to sizes of at least $\sim 1$~mm for the 
vast majority of the disks.   
Within the relatively small samples  investigated so far, 
the distribution of spectral indices is consistent with 
being the same for nearly all the regions probed so far.
The general picture that is emerging from this comparison is that
dust appears to quickly grow to large sizes, but then it needs to be retained 
in the
disk for a relatively long time, comparable to the disk lifetime.
The only region where there may be a hint for possibly different 
distribution of spectral index values is Chamaeleon, 
where \citet{2012MNRAS.425.3137U} derived a range of $\beta$ values 
between 0.9 and 1.8 for 8 disks. Chamaeleon is among the oldest 
regions in the sample \citep[albeit still young with an
estimated median age of $\sim 2$~Ma][]{2007ApJS..173..104L}. It is possible that
the different values of $\alpha$ in Chamaeleon could be an indication for
a time evolution 
of the grain size distribution, with a loss of mm/cm sized pebbles 
relative to smaller grains.  
This suggestion will be tested when 
statistically significant samples in younger and older star forming regions 
are observed with ALMA.

\medskip

Following \citet{birnstiel:2010} and \citet{2012A&A...538A.114P}, we show 
in Fig.~\ref{fig:photometry} the prediction of global grain evolution models in 
disks. \citet{birnstiel:2010} found that the measured 1.1--3~mm spectral indices 
can be well reproduced by models with reasonable values for parameters 
regulating grain fragmentation, gas turbulence and disk structure. However, 
these models are able to explain only the upper envelope of the measured
fluxes (i.e. the most massive disks in the sample). There is a large population
of disks that are difficult to reconcile with the model predictions:
those with low millimeter flux and low spectral index (low-mass disks
containing a substantial amount of large grains). This discrepancy cannot 
be solved by simply reducing the mass of the disk models, as disks with 
lower surface densities would hardly grow grains \citep[][]{birnstiel:2010},
as can be seen by the fragmentation and drift limited growth in Equations~\ref{eq:theo:a_frag} and~\ref{eq:theo:a_drift},
which show that the maximum grain size depend on the dust and gas surface density, respectively.

\medskip

\citet{2012A&A...538A.114P} investigated the effect of time evolution on 
these modeling results, finding that while the radial drift process would 
progressively reduce the disk mass, evolving over a few Ma the models to lower
1~mm fluxes, the drift and fragmentation processes will more
efficiently remove the large grains from the disk, resulting in a 
steep increase of $\alpha$ which would not be consistent with the
observations. \citet{2012A&A...538A.114P} showed that the drift of large 
particles needs to be slowed down, but not halted completely, in order to explain the observed distribution in 
a framework of disk evolution. In this context, it is important to point out 
that no correlation has so far been found between individual stellar 
ages and $\alpha$ \citep[e.g.][]{2010A&A...521A..66R}.

\medskip

Several mechanisms have been proposed to slow down the radial drift of
large grains. For example in MHD 
simulations of disks with zonal flows
\citep{Johansen:2005p8425,2009ApJ...697.1269J,2011ApJ...736...85U}, 
the pressure field in the disk would 
be modified and this may be a viable mechanism to create local pressure
maxima that could efficiently trap large grains. 
Another possibility 
that has been explored by some authors is that very large grains are 
injected very early in the outer disk, then their migration is impaired 
as they are decoupled from the gas \citep[e.g.][]{Laibe:2012p18591}.
This scenario would require the formation of very large (cm-size) grains 
in cores and protostellar envelopes before disk formation, or an extremely
efficient growth in very massive (young) disks. These hypothesis have
not yet been modeled in detail to explore their feasibility.

\medskip

An important prediction of models of dust evolution in gaseous disks 
is that grains are expected to grow to a maximum size that depends on 
the local density of gas and dust \citep[e.g.][]{2009A&A...503L...5B}. 
Therefore, larger values for the mm spectral indices 
should be expected for disks with lower flux at 1~mm, i.e. lower mass 
in dust and presumably lower densities as well. An extreme example of 
such systems could be disks around young brown dwarfs, if they are 
sufficiently large that the surface density is low. Initial measurements
of these systems have confirmed the presence of large grains and 
relatively large radii \citep{2012ApJ...761L..20R,2013ApJ...764L..27R}.
In spite of the very low millimeter fluxes and estimated low dust mass,
the mm spectral indices measured for the disks of $\rho-$Oph 
102 and 2M0444$+$2512 are as low as for the bulk of the T Tauri stars, 
taken together with the information on the disk spatial 
extent from the interferometric observations
\citet{2012ApJ...761L..20R,2013ApJ...764L..27R} constrain 
a value of $\beta \approx 0.5$ for both these brown dwarf disks.
 
\medskip

\citet{2013A&A...554A..95P} and \citet{2013ApJ...774L...4M} have investigated 
in detail the grain growth process in brown dwarf disks, showing that
it is indeed possible to have grain growth and explain the  
values of millimeter flux and spectral index, at least 
in the conditions derived for the disks that have been observed so far.
A more serious problem is represented by the radial drift. To slow down the
radial motion of the grains \citet{2013A&A...554A..95P} had to assume 
rather extreme gas pressure inhomogeneities. ALMA will soon start to 
offer more observational constraints on larger samples of brown dwarf 
disks, allowing a more thorough test of the grain growth models.

\subsubsection{\textbf{Evolution of sub-mm fluxes for disks in different star forming regions}}
\label{Sec:flux_evolution}


Additional evidence for dust grain growth comes from (sub-)millimeter
continuum surveys of star forming regions with different ages.
The key finding is that the disk millimeter luminosity distribution
declines rapidly with time, much quicker than the infrared fraction,
such that very few disks are detected at all in regions older than a few Ma.

The Taurus and Ophiuchus clouds host two of the best studied,
nearby, low mass star forming regions and provide a benchmark
for comparisons with other regions.
Each contains about 200 Class II sources that have been
well characterized at infrared wavelengths through Spitzer surveys
\citep{2009ApJS..181..321E,2010ApJS..186..111L}.
Almost all these sources have been observed at (sub-)millimeter
wavelengths in Taurus \citep[][and references therein]{2013ApJ...771..129A}
and a large survey was carried out in Ophiuchus by \citet{2007ApJ...671.1800A}.
Both regions are relatively young, in the sense of having an infrared disk
fraction, $f_{\rm disk} = N_{\rm disk}/N_{\rm tot} \geq 60$\% and the
disk millimeter distributions are broadly similar, lognormal with a
mean flux density mean $F(1.3\,{\rm mm})=4$\,mJy and standard deviation
0.9 dex \citep{2013ApJ...771..129A}.

However, only a handful of disks are detected at millimeter wavelengths
in more evolved regions
such as Upper Sco \citep[$f_{\rm disk} = 19$\%,][]{2012ApJ...745...23M}
and $\sigma$\,Ori \citep[$f_{\rm disk} = 27$\%][]{2013MNRAS.435.1671W}.
Statistical comparisons must take into account not only the varying
survey depths of course, but also the stellar properties of the samples
as disk masses depend on both stellar binarity and mass \citep{2013ApJ...771..129A}.
The results of a Monte-Carlo sampling technique to allow for these
effects 
demonstrates that infrared Class II
disk millimeter luminosities decrease significantly as regions age
and $f_{\rm disk}$ decreases \citep{2013MNRAS.435.1671W}.

The precise ages (and age spreads) of the compared regions are not well
known but they are all young enough, $\ll 10$\,Myr,
that planet formation should be ongoing.
Therefore, the decrease in millimeter luminosity is attributable more to
a decrease in the emitting surface area per unit mass, i.e., grain growth,
than a decrease in the solid mass \citep{2010MNRAS.407.1981G}.
The relative age differences of Upper Sco and $\sigma$\,Ori with respect to
Taurus are better constrained and are $\sim 3-5$\,Myr.  This is, therefore,
an upper limit to the typical timescale on which most of the solid mass in
disks is locked up into particles greater than about a millimeter in size.

\subsection{\textbf{Resolved Images at Millimeter/Radio Wavelengths}}
\label{resolved_images}

%
%
\label{Sec:Resolved}

\noindent The theoretical models for the evolution of solid particles in 
protoplanetary disks that were described in \S\ref{Sec:Theo} make 
several  physical predictions that should have direct 
observational consequences: (1) settling, growth and 
inward drift conspire to produce a size-sorted vertical and radial 
distribution of solids, 
such that larger particles are preferentially more concentrated in the disk
mid-plane and near the host 
star; (2) drift alone should substantially increase the gas-to-dust mass 
ratio at large disk radii (especially for mm/cm-sized particles); and (3) dust 
transport and fragmentation processes imply that the growth of solids to planetesimals
has to happen in spatially confined regions of the disks.  With their 
high sensitivity to cool gas and dust emission at a wide range of angular 
scales, observations with mm/radio-wavelength interferometers are uniquely 
suited for an empirical investigation of these physical effects.  
Ultimately, such data can be used to benchmark the theoretical models, and then 
provide observational constraints on their key input parameters (e.g. 
turbulence, particle properties, growth timescales, etc.; see \S\ref{Sec:Lab} 
and~\ref{Sec:Theo}).  

\subsubsection{Constraints on large scale radial variation of dust properties in disks}

In the context of the dust continuum emission, we already highlighted in 
\S\ref{Sec:sub-mm_phot} how the disk-averaged mm/radio ``color" -- typically parameterized in 
terms of the spectral index of the dust opacity, $\beta$ --  provides a global 
view of the overall level of particle growth in the disk.  
While this is a useful and efficient 
approach to study the demographics of large surveys,
it cannot tell us about the expected strong spatial variations of the 
dust properties in individual disks 
\citep[e.g.][]{birnstiel:2010}.  A more useful technique would be to map out 
the spatial dependence of the mm/radio colors, $\beta(r)$ or its equivalent, by 
resolving the continuum emission at a range of observing wavelengths.  The 
underlying principle behind this technique rests on the (well justified) 
assumption that the continuum emission at sufficiently long wavelengths is 
optically thin, so that the surface brightness profile scales like
\begin{equation}
I_{\nu}(r) \propto \kappa_{\nu} \, B_{\nu}(T) \, \zeta \, \Sigma_g,
\label{eq:surf}
\end{equation}
where $\kappa_{\nu}$ is the opacity spectrum, $B_{\nu}(T)$ is the Planck 
function at the local temperature, $\zeta$ is the inverse of the gas-to-dust mass ratio, and 
$\Sigma_g$ is the gas surface density profile -- each of which is thought to 
vary spatially in a given disk.  

\medskip

The spectral behavior in Eq.~(\ref{eq:surf}) has been exploited to interpret 
multi-wavelength resolved continuum images in two basic approaches.  The first 
approach acknowledges that the forward-modeling problem is quite difficult, 
since we do not really have an a priori parametric model for the spatial 
variations of $\kappa_{\nu}$, $T$, or $\Sigma_d$ (where the dust surface 
density profile, $\Sigma_d = \zeta \, \Sigma_g$).  \citet{isella:2010} reasoned
that, for a suitable assumption or model of $T(r)$, the spectral gradient of 
the surface brightness profile itself should provide an empirical measurement 
of the resolved mm/radio color regardless of $\Sigma_d$.  In essence, 
parametric fits for the optical depth profiles at each individual wavelength, 
$\tau_{\nu}(r) \approx \kappa_{\nu} \Sigma_d$, can be converted into an opacity 
index profile,
\begin{equation}
\beta(r) = \frac{\partial \log{\kappa_{\nu}(r)}}{\partial \log{\nu}} \approx \frac{\partial \log{\tau_{\nu}(r)}}{\partial \log{\nu}}.  
\label{eq:isella}
\end{equation}
Although the initial studies that adopted this approach had insufficient data 
to conclusively argue for a non-constant $\beta(r)$, they did establish an 
empirically motivated technique that provides a straightforward means of 
mapping mm/radio colors \citep[e.g.][]{isella:2010,banzatti:2011}.  Put 
simply, this approach allows one to reconstruct the $\beta(r)$ required to 
reconcile seemingly discrepant continuum emission structures at different 
observing wavelengths.  

\medskip

A slight variation on this approach is to adopt a more typical forward-modeling 
technique, where an assumption is made for a parametric formulation for both 
$\kappa_{\nu}$ and $\Sigma_d$.  For example, \citet{guilloteau:2011} explored 
their dual-wavelength observations of Taurus disks with power-law and 
step-function $\beta(r)$ profiles, and argued that a spectral index that 
increases with disk radius provides a substantially improved fit quality 
compared to a global, constant index.  In a recent refinement of the 
\citet{banzatti:2011} work, \citet{2013A&A...558A..64T} have developed a more 
physically motivated prescription for $\kappa_{\nu}(r)$ (parameterized as a 
function of the local grain size distribution) that clearly calls for an 
increasing $\beta(r)$ in the disk around CQ Tau.  

\medskip

\begin{figure*}[th]
\includegraphics[width=8.5cm]{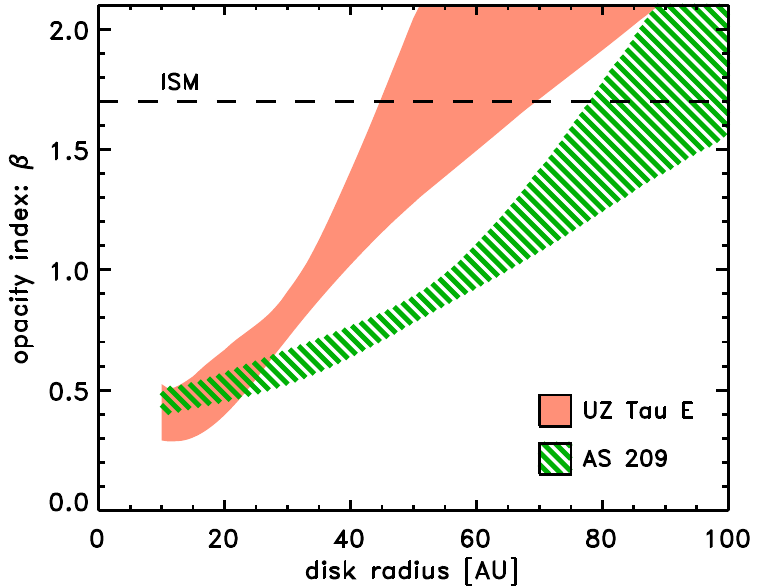}
\includegraphics[width=8.5cm]{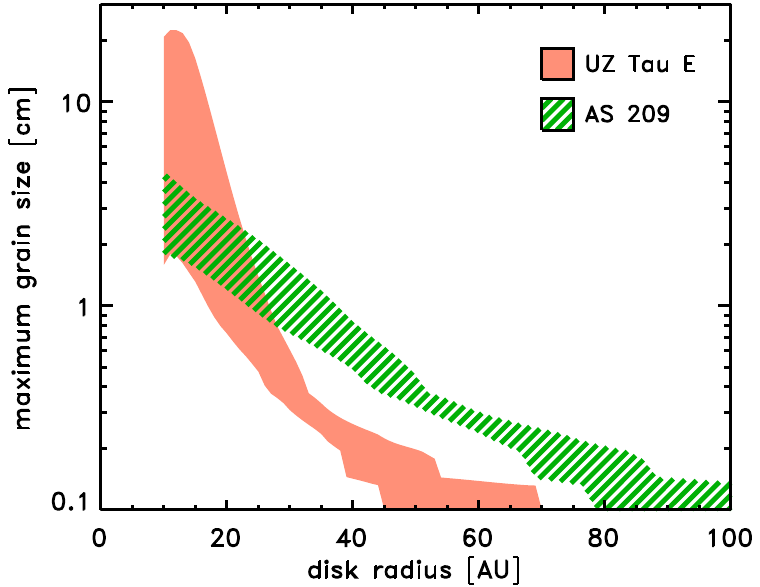}
\caption{
Left panel: confidence 
ranges for the profiles of $\beta$ as a function of radius for the two disks 
surrounding the young stars AS~209 and UZ~Tau~E, as derived from sub-millimeter
through centimeter wave high signal to noise and high angular resolution 
interferometric observations 
\citep[][ {\it Harris, priv. comm.}]{perez:2012}.
Right panel: values of a$_{max}$ as a function of radius for the AS~209 and UZ~Tau~E
disks derived with our reference dust model. The dashed and continuos lines
show the fragmentation and radial drift limit, respectively, for a representative
global model of dust evolution in disks (\S~\ref{Sec:Theo}).
}
\label{fig:evla_beta}
\end{figure*}

In these initial studies, the fundamental technical obstacle was really the 
limited wavelength range over which sensitive, resolved continuum measurements 
were available (typically $\lambda = 1.3$--2.7\,mm).  An extension of this work 
to centimeter wavelengths offers a substantially increased leverage on 
determining $\beta(r)$ and uniquely probes the largest detectable solid 
particles, while also overcoming the systematic uncertainties related to the 
absolute amplitude calibration for individual datasets.  In two recent studies 
of the disks around AS 209 \citep{perez:2012} and UZ Tau E 
({\it Harris, priv. comm.}),
resolved continuum emission at $\lambda \approx 9$\,mm from the upgraded Jansky 
Very Large Array (VLA) was folded into the analysis to provide robust evidence 
for significant increases in their $\beta(r)$ profiles (as shown together in 
Figure \ref{fig:evla_beta}).  The analysis of the $\beta(r)$ profiles,
as inferred from the optical depth profiles, for these two 
objects suggest at least an order of magnitude increase in particle 
sizes when moving in from $r \approx 100$\,AU to $\sim$10\,AU scales.

\medskip

Taken together, these multi-wavelength interferometric dust continuum 
measurements reveal a fundamental, and apparently general, observational 
feature in young protoplanetary disks: the size of the dust emission 
region is 
anti-correlated with the observing wavelength.  The sense of that relationship 
is in excellent agreement with the predictions of theoretical models for the 
evolution of disk solids, where the larger particles that emit more efficiently 
at longer wavelengths are concentrated at small disk radii due to the combined 
effects of growth and drift.  The quantitative characterization of this feature 
in individual disks is just getting started, but the promise of a new 
opportunity to leverage current observing facilities to constrain planetesimal 
formation in action is  exciting.  However, there is a downside: it 
is now clear that resolved observations at a single mm/radio wavelength are not 
sufficient to constrain fundamental parameters related to the dust density 
structure.  Given the lingering uncertainty on an appropriate general 
parameterization for $\Sigma_d$, it is worthwhile to point out that the 
optimized mechanics of the modeling approach for these resolved multi-wavelength 
continuum data are still in a stage of active development.  Yet, as more data 
becomes available, rapid advances are expected from both the theoretical 
and observational communities.

\medskip

The behavior of the multi-wavelength dust continuum emission makes a strong 
case for the spatial variation of $\kappa_{\nu}$ induced by the growth and 
migration of disk solids.  But, as mentioned at the start of this section, 
those same physical processes should also produce a complementary discrepancy 
between the spatial distribution of the gas and dust phases in a disk: the dust 
should be preferentially more concentrated toward the stellar host, and  
$\zeta$ should decrease dramatically with radius as the gas-to-dust ratio
increases in the outer disk.  
The fundamental problem is that the derived dust densities are very uncertain, 
and, even more severe, we do not yet really understand how to 
measure gas densities in these disks, so we cannot infer $\zeta(r)$ 
with sufficient quantitative reliability.  

\medskip

However, an approach that relies on a comparison between the spatial extents of 
molecular line and dust continuum emission can provide an indirect constraint 
on the spatial variation of $\zeta(r)$, even if its normalization remains 
uncertain.  It has been recognized for some time that the sizes of the gas 
disks traced by optically thick CO line emission appear systematically larger 
than their optically thin dust continuum 
\citep[e.g.][]{pietu:2007,isella:2007}.  Although previous work suggested that 
this may be an artificial feature caused by optical depth effects or a 
misleading density model \citep{hughes:2008}, the CO--dust size discrepancies 
have persisted with more sophisticated models and improved sensitivity.  The 
evidence for a decreasing $\zeta(r)$ profile is derived through a process of 
logical negation.  First, the dust structure of an individual disk is modeled, 
based on radiative transfer calculations that match the broadband SED and one 
or more resolved continuum images.  Then, a gas structure model is calculated 
assuming a spatially invariant $\zeta$.  A comparison of the corresponding 
synthetic CO emission model and the data invariably demonstrates that the model 
produces line emission that is much too compact. Small grains are found to be
well mixed with the gas out to very large radii, e.g. from scattered light 
observations (see \S~\ref{Sec:scattir}). These findings imply that the large 
grains are confined to a smaller region of the disk, as compared to the gas
and small dust grains. It is reasonable to deduce that a combination of grain 
size radial segregation and a decreasing $\zeta(r)$ would reconcile the line and 
continuum data.

\medskip

As of this writing, only 5 disks have been modeled in this manner to indirectly 
infer a decreasing $\zeta(r)$, around IM~Lup \citep{panic:2008}, TW~Hya 
\citep{andrews:2012}, LkCa~15 \citep{isella:2012}, V4046~Sgr 
\citep{2013ApJ...775..136R}, HD~163296 \citep{2013A&A...557A.133D}.  
Ultimately, an empirical measurement of the 
CO--dust size discrepancy could be combined with the $\beta(r)$ profile 
inferences to help better constrain drift rates and how the migration of solids 
depends on the local disk conditions.

\subsubsection{Constraints on vertical stratification of dust properties in disks}

In \S~\ref{Sec:scattir} we discussed the important constraints that 
infrared and scattered light observations provide on the grain populations
in the disk atmosphere and on the dust settling \citep[an extensive discussion
of these can also be found in][]{2007prpl.conf..767N}. The initial attempts to
relate the grain properties in the disk atmosphere, as derived from mid 
infrared spectroscopy, with the properties 
on the mid-plane, as derived from (sub-)millimeter photometry, have met very
limits success, even when considering large samples 
\citep[e.g.][]{2009A&A...495..869L,2010A&A...521A..66R,2010ApJ...721..431J,2012MNRAS.425.3137U}.
While models do predict a relationship between the growth level 
on the mid-plane and the atmosphere of the disks, the atmosphere grains should reach a steady 
state balance between growth and fragmentation while mid-plane dust is 
still growing, in addition another likely cause of the difficulty
in recovering a correlation between the dust properties observationally 
is most likely linked to the 
vastly different regions of the disks that are currently sampled in the 
two wavelength regimes \citep[e.g.][]{2007prpl.conf..767N}. A meaningful 
comparison will only be possible with data sampling the same regions of the disk,
which will require substantial improvement of the mid-infrared observations,
both in terms of sensitivity and angular resolution, which will become 
possible with the 40-m class ELTs.

\medskip

Better results are being obtained by comparing the dust properties as derived from high
angular resolution optical-infrared images and millimeter dust emission maps.
Constraints on the dust settling in the upper layers of the disk atmosphere 
have been discussed extensively in the last decade 
\citep[\S\ref{Sec:scattir} and, e.g.][]{2003ApJ...588..373W,2008ApJ...674L.101W,2012A&ARv..20...52W,2012A&A...546A...7L},
so far, the main limitation to combined multi-wavelength studies that
could constrain dust properties across the whole disk scale height
has been the limited sensitivity and angular resolution of (sub-)millimeter
observations.
Some initial studies in this direction are promising, although these are currently
limited to very few objects and with very favorable viewing geometries
\citep[e.g.][]{2013A&A...553A..69G}.
It is expected that ALMA will provide the angular resolution and sensitivity
to significantly progress in this area, although, obviously, the best constraints
will continue to be possible only for disks with very favorable geometries
\citep{2013MNRAS.431.1573B}.

\subsubsection{Constraints on small scale variation of dust properties and dust trapping}

One key development that is expected in the near future is the 
direct observation of the expected small scale structures within the 
disk where solids will be able to overcome the many growth barriers
discussed in \S\ref{Sec:Lab} and~\ref{Sec:Theo} and reach the planetesimal 
stage (see also the chapter by Anders et al.). All of the proposed 
solutions to the long standing m-size barrier problem in planet formation
\citep[][ and all subsequent evolutions]{1977MNRAS.180...57W} involve
the confinement of the largest particles in localized areas of the disk
where they can locally overcome the barriers on their way to become planetesimals.
The typical dimensions of these large grains ``traps'', as they are normally 
referred to, is of the order of the local scale height of the disk or less.
Several studies have simulated in detail the observability of these features in 
young protoplanetary disks with ALMA showing that there are good prospects to
reveal them both in the gas and/or dust emission
\citep[e.g.][]{2010MNRAS.407..181C,2012A&A...538A.114P,2013A&A...554A..95P,2013MNRAS.433.2064D}.

\medskip

A very recent and extremely encouraging development in this direction is the 
detection with ALMA of a large dust trap in a transition disk by 
\citet{2013Sci...340.1199V}. These authors revealed clearly the segregation of 
large dust particles in the outer regions of the disk caused by the radial
and azimuthal inhomogeneity in the gas density induced by the likely presence
of a planetary companion within the disk inner hole. These data provide a 
direct observational support to the simulations of dust trapping in 
transitional disks as discussed by \citet{2012A&A...545A..81P}. While the 
result of \citet{2013Sci...340.1199V} provide a direct proof of the dust 
trapping concept, the characteristics of the observed system are such that
the trap may support the formation of an analog to the Solar System Kuiper-Belt,
but cannot obviously explain the formation of planets in that system, as the presence of 
a planet is a requirement for the formation of the trap. Nevertheless, the 
result illustrates how we can expect to constrain dust trapping in disks 
with ALMA in the future. 

\medskip 

Sensitive and high angular resolution observations with ALMA will also very soon
allow us to understand the role of snowlines in the evolution of solids in
protoplanetary disks. Grains that are migrating inward across a snowline 
will loose part of their icy mantles and models predict that the cycle of
sublimation and condensation will allow efficient growth and trapping across the
snowline \citep[e.g.][]{Ros:2013p20760}.
\citet{2013A&A...557A.132M} and \citet{2013Sci...341..630Q} resolved spatially
the CO snowline in HD~163296 and TW~Hya, paving the road for an investigation
of the role of this particular snowline in the evolution of solids.

\section{\textbf{SUMMARY AND OUTLOOK}}
\label{Sec:Summary}

Global models of grain evolution in disks, constrained by the results
of laboratory and numerical calculations of grain and aggregate collisions,
predict that in the conditions of protoplanetary disks grains should 
very rapidly grow to centimeter sizes. The growth process is not without 
difficulties, with numerous ``barriers'' to overcome. At this time,
the most critical problem seems to be the drift-fragmentation ``barrier'' 
caused by the large differential radial speeds of grains of different 
sizes induced by the aerodynamical drag. While the level of growth 
predicted by models is consistent with observational data, the time evolution 
is too fast unless large grains radial drift motion is slowed down in
pressure ``traps''. Several ideas of plausible mechanisms to produce these
traps exist: from disk instabilities to snowlines. The predictions should now
be tested observationally and ALMA will offer a unique opportunity to do this in the 
coming years thanks to its superb angular resolution and sensitivity.

\medskip

Observations show that the majority of disks around isolated
pre-main sequence stars in the nearby star forming regions contain 
a significant amount of grains grown to at least millimeter sizes. 
No clear evolutionary trends have emerged so far, although some 
indirect evidence of disk aging is starting to emerge also from millimeter
observations. This finding strongly supports the notion that grain growth 
is a common and very fast process in disks, possibly occurring in the 
earliest phases of disk formation. The other implication is that the
large grains are either kept in the disk mid-plane  or they 
are continuously reformed for several Ma. This is required to explain the 
presence of these dust aggregates in disks around pre-main sequence stars and also
consistent with cosmochemical evidence in our own Solar System.

\medskip

Multi-wavelength radially resolved observations of grain properties, which 
were just barely becoming possible at the time of the last Protostars and Planets 
conference, are now providing a wealth of new constraints to the models.
The overall radial segregation of dust aggregates of different sizes predicted by models
is in general agreement with observations, however, the fast draining of solids
towards the inner disk predicted by models cannot reproduce the observations. 
A mechanism to slow down the radial drift is definitely required. 

\medskip

While the interpretation of the millimeter observations in terms of grain
growth appears to be solid, the values of the dust opacities for the 
dust aggregates are still very uncertain. New computations or laboratory 
measurements for a range of grain composition and structure would represent
a major step to put on very solid grounds the study of dust evolution in disks.
Similarly, a major step in the understanding of the physical processes 
of grain growth in disks will be the extension of the grain-grain laboratory
experiments to icy dust particles. These experiments are now been carried out 
and the initial results are expected in the coming years.

\medskip

Armed with better constraints on the grain-grain collision outcomes, the global 
models of dust evolution in disks will also have to be advanced to the next 
stage. This will have to include a proper account of the gas and dust 
coevolution and the extension of the models to two and three dimensions, 
to account for vertical transport and azimuthal inhomogeneities, which are now 
routinely observed with ALMA.

\medskip

On the observational side new and upgraded 
observing facilities at millimeter and centimeter wavelengths are offering
an unprecedented opportunity to expand the detailed studies to faint 
objects and to resolve the detailed radial and vertical structure of the 
grain properties. In the coming years we expect that it will be possible 
to put  strong constraints on the evolutionary timescale for the 
dust on the disk mid-plane and explore this process as a function of the 
central host star parameters and as a function of environment. 
The study of the dust and gas small scale structures in the disk, as well 
as their global distribution, will most likely allow us to solve the 
long standing problem of radial drift.

\textbf{ Acknowledgments.} We thank M.~Bizzarro, P.~Caselli, C.~Chandler, C.~Dullemond, A.~Dutrey, Th.~Henning, A.~Johansen, C.~Ormel, R.~Waters, and especially the referee, S. Okuzumi, for valuable comments and discussions on the content of this chapter. 
We thank F.~Windmark for providing data for Fig.~\ref{fig:lab}.
The work reported here has been partly supported over the years by grants to INAF-Osservatorio Astrofisico di Arcetri
from the MIUR, PRIN-INAF and ASI. Extended support from the Munich-IMPRS as well as ESO Studentship/Internship, DGDF 
and Scientific Visitor programmes in the period 2009-2013 is gratefully acknowledged. 
S.~Andrews and T.~Birnstiel acknowledge support from NASA Origins of Solar Systems grant NNX12AJ04G. 
A.~Isella and J. M.~Carpenter acknowledge support from NSF award AST-1109334.
J. P. Williams acknowledges support from NSF award AST-1208911.

\bigskip

\bibliographystyle{ppvi_lim1.bst}
\bibliography{ltesti_ppvi}

\begin{thebibliography}{255}
\parskip=0pt \itemsep=0pt \small \baselineskip=11pt
\providecommand{\natexlab}[1]{#1}

\bibitem[\protect\astroncite{\emph{{{\'A}brah{\'a}m}
  et~al.}}{2009}]{2009Natur.459..224A}
{{\'A}brah{\'a}m} P. et~al. (2009) \emph{\nat}, \emph{459}, 224.

\bibitem[\protect\astroncite{\emph{{Adachi}
  et~al.}}{1976}]{1976PThPh..56.1756A}
{Adachi} I. et~al. (1976) \emph{Progress of Theoretical Physics}, \emph{56},
  1756.

\bibitem[\protect\astroncite{\emph{{Adams} et~al.}}{1990}]{1990ApJ...357..606A}
{Adams} F.~C. et~al. (1990) \emph{\apj}, \emph{357}, 606.

\bibitem[\protect\astroncite{\emph{{Andrews} and
  {Williams}}}{2005}]{2005ApJ...631.1134A}
{Andrews} S.~M. and {Williams} J.~P. (2005) \emph{\apj}, \emph{631}, 1134.

\bibitem[\protect\astroncite{\emph{{Andrews} and
  {Williams}}}{2007}]{2007ApJ...671.1800A}
{Andrews} S.~M. and {Williams} J.~P. (2007) \emph{\apj}, \emph{671}, 1800.

\bibitem[\protect\astroncite{\emph{{Andrews} et~al.}}{2012}]{andrews:2012}
{Andrews} S.~M. et~al. (2012) \emph{\apj}, \emph{744}, 162.

\bibitem[\protect\astroncite{\emph{{Andrews}
  et~al.}}{2013}]{2013ApJ...771..129A}
{Andrews} S.~M. et~al. (2013) \emph{\apj}, \emph{771}, 129.

\bibitem[\protect\astroncite{\emph{{Ardila}
  et~al.}}{2007}]{2007ApJ...665..512A}
{Ardila} D.~R. et~al. (2007) \emph{\apj}, \emph{665}, 512.

\bibitem[\protect\astroncite{\emph{{Armitage}}}{2010}]{2010apf..book.....A}
{Armitage} P.~J. (2010) \emph{{Astrophysics of Planet Formation}}, Cambridge
  University Press.

\bibitem[\protect\astroncite{\emph{Ataiee et~al.}}{2013}]{Ataiee:2013p20827}
Ataiee S. et~al. (2013) \emph{A{\&}A}, \emph{553}, L3.

\bibitem[\protect\astroncite{\emph{Bai and Stone}}{2010}]{Bai:2010p15702}
Bai X.-N. and Stone J.~M. (2010) \emph{ApJ}, \emph{722}, 1437.

\bibitem[\protect\astroncite{\emph{{Banzatti} et~al.}}{2011}]{banzatti:2011}
{Banzatti} A. et~al. (2011) \emph{\aap}, \emph{525}, A12.

\bibitem[\protect\astroncite{\emph{{Barge} and
  {Sommeria}}}{1995}]{1995A&A...295L...1B}
{Barge} P. and {Sommeria} J. (1995) \emph{\aap}, \emph{295}, L1.

\bibitem[\protect\astroncite{\emph{{Beckwith} and
  {Sargent}}}{1991}]{1991ApJ...381..250B}
{Beckwith} S.~V.~W. and {Sargent} A.~I. (1991) \emph{ApJ}, \emph{381}, 250.

\bibitem[\protect\astroncite{\emph{Beckwith
  et~al.}}{2000}]{Beckwith:2000p20930}
Beckwith S. V.~W. et~al. (2000) \emph{PP IV}, p. 533.

\bibitem[\protect\astroncite{\emph{{Birnstiel}
  et~al.}}{2009}]{2009A&A...503L...5B}
{Birnstiel} T. et~al. (2009) \emph{\aap}, \emph{503}, L5.

\bibitem[\protect\astroncite{\emph{{Birnstiel}
  et~al.}}{2010{\natexlab{a}}}]{2010A&A...513A..79B}
{Birnstiel} T. et~al. (2010{\natexlab{a}}) \emph{\aap}, \emph{513}, A79.

\bibitem[\protect\astroncite{\emph{{Birnstiel}
  et~al.}}{2010{\natexlab{b}}}]{birnstiel:2010}
{Birnstiel} T. et~al. (2010{\natexlab{b}}) \emph{\aap}, \emph{516}, L14.

\bibitem[\protect\astroncite{\emph{{Birnstiel}
  et~al.}}{2011}]{2011A&A...525A..11B}
{Birnstiel} T. et~al. (2011) \emph{\aap}, \emph{525}, A11.

\bibitem[\protect\astroncite{\emph{{Birnstiel}
  et~al.}}{2012}]{2012A&A...539A.148B}
{Birnstiel} T. et~al. (2012) \emph{\aap}, \emph{539}, A148.

\bibitem[\protect\astroncite{\emph{{Birnstiel}
  et~al.}}{2013}]{2013A&A...550L...8B}
{Birnstiel} T. et~al. (2013) \emph{\aap}, \emph{550}, L8.

\bibitem[\protect\astroncite{\emph{{Blum}}}{2006}]{2006AdPhy..55..881B}
{Blum} J. (2006) \emph{Advances in Physics}, \emph{55}, 881.

\bibitem[\protect\astroncite{\emph{{Blum} and
  {Wurm}}}{2000}]{2000Icar..143..138B}
{Blum} J. and {Wurm} G. (2000) \emph{\icarus}, \emph{143}, 138.

\bibitem[\protect\astroncite{\emph{{Blum} and
  {Wurm}}}{2008}]{2008ARA&A..46...21B}
{Blum} J. and {Wurm} G. (2008) \emph{\araa}, \emph{46}, 21.

\bibitem[\protect\astroncite{\emph{{Blum} et~al.}}{2000}]{2000PhRvL..85.2426B}
{Blum} J. et~al. (2000) \emph{Physical Review Letters}, \emph{85}, 2426.

\bibitem[\protect\astroncite{\emph{Bockel{\'e}e-Morvan
  et~al.}}{2002}]{BockeleeMorvan:2002p20828}
Bockel{\'e}e-Morvan D. et~al. (2002) \emph{Astronomy and Astrophysics},
  \emph{384}, 1107.

\bibitem[\protect\astroncite{\emph{{Boehler}
  et~al.}}{2013}]{2013MNRAS.431.1573B}
{Boehler} Y. et~al. (2013) \emph{\mnras}, \emph{431}, 1573.

\bibitem[\protect\astroncite{\emph{{Bohren} and
  {Huffman}}}{1983}]{1983asls.book.....B}
{Bohren} C.~F. and {Huffman} D.~R. (1983) \emph{{Absorption and scattering of
  light by small particles}}, Wiley.

\bibitem[\protect\astroncite{\emph{{Bouwman}
  et~al.}}{2001}]{2001A&A...375..950B}
{Bouwman} J. et~al. (2001) \emph{\aap}, \emph{375}, 950.

\bibitem[\protect\astroncite{\emph{Brauer et~al.}}{2007}]{Brauer:2007p232}
Brauer F. et~al. (2007) \emph{A{\&}A}, \emph{469}, 1169.

\bibitem[\protect\astroncite{\emph{{Brauer}
  et~al.}}{2008{\natexlab{a}}}]{2008A&A...480..859B}
{Brauer} F. et~al. (2008{\natexlab{a}}) \emph{\aap}, \emph{480}, 859.

\bibitem[\protect\astroncite{\emph{{Brauer}
  et~al.}}{2008{\natexlab{b}}}]{2008A&A...487L...1B}
{Brauer} F. et~al. (2008{\natexlab{b}}) \emph{\aap}, \emph{487}, L1.

\bibitem[\protect\astroncite{\emph{Carballido
  et~al.}}{2006}]{Carballido:2006p18250}
Carballido A. et~al. (2006) \emph{MNRAS}, \emph{373}, 1633.

\bibitem[\protect\astroncite{\emph{{Carballido}
  et~al.}}{2010}]{2010MNRAS.405.2339C}
{Carballido} A. et~al. (2010) \emph{\mnras}, \emph{405}, 2339.

\bibitem[\protect\astroncite{\emph{{Chatterjee} and
  {Tan}}}{2014}]{2014ApJ...780...53C}
{Chatterjee} S. and {Tan} J.~C. (2014) \emph{\apj}, \emph{780}, 53.

\bibitem[\protect\astroncite{\emph{Chiang and
  Laughlin}}{2013}]{Chiang:2013p20955}
Chiang E. and Laughlin G. (2013) \emph{MNRAS}, \emph{431}, 3444.

\bibitem[\protect\astroncite{\emph{{Chiang} and
  {Goldreich}}}{1997}]{1997ApJ...490..368C}
{Chiang} E.~I. and {Goldreich} P. (1997) \emph{\apj}, \emph{490}, 368.

\bibitem[\protect\astroncite{\emph{Chiang et~al.}}{2001}]{Chiang:2001p7463}
Chiang E.~I. et~al. (2001) \emph{ApJ}, \emph{547}, 1077.

\bibitem[\protect\astroncite{\emph{{Chiang}
  et~al.}}{2012}]{2012ApJ...756..168C}
{Chiang} H.-F. et~al. (2012) \emph{\apj}, \emph{756}, 168.

\bibitem[\protect\astroncite{\emph{{Chokshi}
  et~al.}}{1993}]{1993ApJ...407..806C}
{Chokshi} A. et~al. (1993) \emph{\apj}, \emph{407}, 806.

\bibitem[\protect\astroncite{\emph{Ciesla}}{2009}]{Ciesla:2009p10132}
Ciesla F.~J. (2009) \emph{Icarus}, \emph{200}, 655.

\bibitem[\protect\astroncite{\emph{{Clampin}
  et~al.}}{2003}]{2003AJ....126..385C}
{Clampin} M. et~al. (2003) \emph{\aj}, \emph{126}, 385.

\bibitem[\protect\astroncite{\emph{{Connelly}
  et~al.}}{2012}]{2012Sci...338..651C}
{Connelly} J.~N. et~al. (2012) \emph{Science}, \emph{338}, 651.

\bibitem[\protect\astroncite{\emph{{Cossins}
  et~al.}}{2010}]{2010MNRAS.407..181C}
{Cossins} P. et~al. (2010) \emph{\mnras}, \emph{407}, 181.

\bibitem[\protect\astroncite{\emph{Cuzzi and Hogan}}{2003}]{Cuzzi:2003p12258}
Cuzzi J.~N. and Hogan R.~C. (2003) \emph{Icarus}, \emph{164}, 127.

\bibitem[\protect\astroncite{\emph{Cuzzi et~al.}}{1993}]{Cuzzi:1993p15730}
Cuzzi J.~N. et~al. (1993) \emph{Icarus}, \emph{106}, 102.

\bibitem[\protect\astroncite{\emph{D'Alessio
  et~al.}}{2001}]{DAlessio:2001p4130}
D'Alessio P. et~al. (2001) \emph{ApJ}, \emph{553}, 321.

\bibitem[\protect\astroncite{\emph{{D'Alessio}
  et~al.}}{2006}]{2006ApJ...638..314D}
{D'Alessio} P. et~al. (2006) \emph{\apj}, \emph{638}, 314.

\bibitem[\protect\astroncite{\emph{{Dauphas} and
  {Chaussidon}}}{2011}]{2011AREPS..39..351D}
{Dauphas} N. and {Chaussidon} M. (2011) \emph{Annual Review of Earth and
  Planetary Sciences}, \emph{39}, 351.

\bibitem[\protect\astroncite{\emph{{de Gregorio-Monsalvo}
  et~al.}}{2013}]{2013A&A...557A.133D}
{de Gregorio-Monsalvo} I. et~al. (2013) \emph{\aap}, \emph{557}, A133.

\bibitem[\protect\astroncite{\emph{{Dominik} and
  {Dullemond}}}{2008}]{2008A&A...491..663D}
{Dominik} C. and {Dullemond} C.~P. (2008) \emph{\aap}, \emph{491}, 663.

\bibitem[\protect\astroncite{\emph{{Dominik} and
  {N{\"u}bold}}}{2002}]{2002Icar..157..173D}
{Dominik} C. and {N{\"u}bold} H. (2002) \emph{\icarus}, \emph{157}, 173.

\bibitem[\protect\astroncite{\emph{{Dominik} and
  {Tielens}}}{1997}]{1997ApJ...480..647D}
{Dominik} C. and {Tielens} A.~G.~G.~M. (1997) \emph{\apj}, \emph{480}, 647.

\bibitem[\protect\astroncite{\emph{Dominik et~al.}}{2007}]{Dominik:2007p1420}
Dominik C. et~al. (2007) \emph{PP V}, p. 783.

\bibitem[\protect\astroncite{\emph{{Douglas}
  et~al.}}{2013}]{2013MNRAS.433.2064D}
{Douglas} T.~A. et~al. (2013) \emph{\mnras}, \emph{433}, 2064.

\bibitem[\protect\astroncite{\emph{{Draine}}}{2006}]{2006ApJ...636.1114D}
{Draine} B.~T. (2006) \emph{ApJ}, \emph{636}, 1114.

\bibitem[\protect\astroncite{\emph{{Dubrulle}
  et~al.}}{1995}]{1995Icar..114..237D}
{Dubrulle} B. et~al. (1995) \emph{\icarus}, \emph{114}, 237.

\bibitem[\protect\astroncite{\emph{{Duch{\^e}ne}
  et~al.}}{2010}]{2010ApJ...712..112D}
{Duch{\^e}ne} G. et~al. (2010) \emph{\apj}, \emph{712}, 112.

\bibitem[\protect\astroncite{\emph{{Dullemond} and
  {Dominik}}}{2004}]{2004A&A...421.1075D}
{Dullemond} C.~P. and {Dominik} C. (2004) \emph{\aap}, \emph{421}, 1075.

\bibitem[\protect\astroncite{\emph{{Dullemond} and
  {Dominik}}}{2005}]{2005A&A...434..971D}
{Dullemond} C.~P. and {Dominik} C. (2005) \emph{\aap}, \emph{434}, 971.

\bibitem[\protect\astroncite{\emph{{Dullemond} and
  {Monnier}}}{2010}]{2010ARA&A..48..205D}
{Dullemond} C.~P. and {Monnier} J.~D. (2010) \emph{\araa}, \emph{48}, 205.

\bibitem[\protect\astroncite{\emph{{Dullemond}
  et~al.}}{2001}]{2001ApJ...560..957D}
{Dullemond} C.~P. et~al. (2001) \emph{\apj}, \emph{560}, 957.

\bibitem[\protect\astroncite{\emph{{Dullemond}
  et~al.}}{2007}]{2007prpl.conf..555D}
{Dullemond} C.~P. et~al. (2007) \emph{Protostars and Planets V}, pp. 555--572.

\bibitem[\protect\astroncite{\emph{{Eisner}
  et~al.}}{2008}]{2008ApJ...683..304E}
{Eisner} J.~A. et~al. (2008) \emph{\apj}, \emph{683}, 304.

\bibitem[\protect\astroncite{\emph{{Evans} et~al.}}{2009}]{2009ApJS..181..321E}
{Evans} II N.~J. et~al. (2009) \emph{\apjs}, \emph{181}, 321.

\bibitem[\protect\astroncite{\emph{{Foster}
  et~al.}}{2013}]{2013MNRAS.428.1606F}
{Foster} J.~B. et~al. (2013) \emph{\mnras}, \emph{428}, 1606.

\bibitem[\protect\astroncite{\emph{Fromang and
  Nelson}}{2005}]{Fromang:2005p20823}
Fromang S. and Nelson R.~P. (2005) \emph{Monthly Notices of the Royal
  Astronomical Society: Letters}, \emph{364}, L81.

\bibitem[\protect\astroncite{\emph{{Fromang} and
  {Nelson}}}{2009}]{2009A&A...496..597F}
{Fromang} S. and {Nelson} R.~P. (2009) \emph{\aap}, \emph{496}, 597.

\bibitem[\protect\astroncite{\emph{{Fromang}
  et~al.}}{2011}]{2011A&A...534A.107F}
{Fromang} S. et~al. (2011) \emph{\aap}, \emph{534}, A107.

\bibitem[\protect\astroncite{\emph{{Fukagawa}
  et~al.}}{2010}]{2010PASJ...62..347F}
{Fukagawa} M. et~al. (2010) \emph{\pasj}, \emph{62}, 347.

\bibitem[\protect\astroncite{\emph{{Furlan}
  et~al.}}{2006}]{2006ApJS..165..568F}
{Furlan} E. et~al. (2006) \emph{\apjs}, \emph{165}, 568.

\bibitem[\protect\astroncite{\emph{{Garaud}}}{2007}]{2007ApJ...671.2091G}
{Garaud} P. (2007) \emph{\apj}, \emph{671}, 2091.

\bibitem[\protect\astroncite{\emph{{Garaud}
  et~al.}}{2013}]{2013ApJ...764..146G}
{Garaud} P. et~al. (2013) \emph{\apj}, \emph{764}, 146.

\bibitem[\protect\astroncite{\emph{Goldreich and
  Ward}}{1973}]{Goldreich:1973p11184}
Goldreich P. and Ward W.~R. (1973) \emph{ApJ}, \emph{183}, 1051.

\bibitem[\protect\astroncite{\emph{{Gr{\"a}fe}
  et~al.}}{2013}]{2013A&A...553A..69G}
{Gr{\"a}fe} C. et~al. (2013) \emph{\aap}, \emph{553}, A69.

\bibitem[\protect\astroncite{\emph{{Greaves} and
  {Rice}}}{2010}]{2010MNRAS.407.1981G}
{Greaves} J.~S. and {Rice} W.~K.~M. (2010) \emph{\mnras}, \emph{407}, 1981.

\bibitem[\protect\astroncite{\emph{{Guilloteau}
  et~al.}}{2011}]{guilloteau:2011}
{Guilloteau} S. et~al. (2011) \emph{\aap}, \emph{529}, A105.

\bibitem[\protect\astroncite{\emph{{Gundlach}
  et~al.}}{2011}]{2011Icar..214..717G}
{Gundlach} B. et~al. (2011) \emph{\icarus}, \emph{214}, 717.

\bibitem[\protect\astroncite{\emph{{G{\"u}ttler}
  et~al.}}{2010}]{2010A&A...513A..56G}
{G{\"u}ttler} C. et~al. (2010) \emph{\aap}, \emph{513}, A56.

\bibitem[\protect\astroncite{\emph{{Hartmann}}}{2009}]{2009apsf.book.....H}
{Hartmann} L. (2009) \emph{{Accretion Processes in Star Formation: Second
  Edition}}, Cambridge University Press.

\bibitem[\protect\astroncite{\emph{Hayashi}}{1981}]{Hayashi:1981p15696}
Hayashi C. (1981) \emph{Prog. Theor. Phys. Suppl.}, \emph{70}, 35.

\bibitem[\protect\astroncite{\emph{{Heim} et~al.}}{1999}]{1999PhRvL..83.3328H}
{Heim} L.-O. et~al. (1999) \emph{Physical Review Letters}, \emph{83}, 3328.

\bibitem[\protect\astroncite{\emph{Hirashita and
  Kuo}}{2011}]{Hirashita:2011p16524}
Hirashita H. and Kuo T.-M. (2011) \emph{MNRAS}, \emph{416}, 1340.

\bibitem[\protect\astroncite{\emph{Hsieh and Gu}}{2012}]{Hsieh:2012p21441}
Hsieh H.-F. and Gu P.-G. (2012) \emph{ApJ}, \emph{760}, 119.

\bibitem[\protect\astroncite{\emph{Hubbard}}{2012}]{Hubbard:2012p20911}
Hubbard A. (2012) \emph{MNRAS}, \emph{426}, 784.

\bibitem[\protect\astroncite{\emph{Hughes and
  Armitage}}{2012}]{Hughes:2012p20952}
Hughes A. L.~H. and Armitage P.~J. (2012) \emph{MNRAS}, \emph{423}, 389.

\bibitem[\protect\astroncite{\emph{{Hughes} et~al.}}{2008}]{hughes:2008}
{Hughes} A.~M. et~al. (2008) \emph{\apj}, \emph{678}, 1119.

\bibitem[\protect\astroncite{\emph{{Isella}
  et~al.}}{2006}]{2006A&A...451..951I}
{Isella} A. et~al. (2006) \emph{\aap}, \emph{451}, 951.

\bibitem[\protect\astroncite{\emph{{Isella} et~al.}}{2007}]{isella:2007}
{Isella} A. et~al. (2007) \emph{\aap}, \emph{469}, 213.

\bibitem[\protect\astroncite{\emph{{Isella} et~al.}}{2010}]{isella:2010}
{Isella} A. et~al. (2010) \emph{\apj}, \emph{714}, 1746.

\bibitem[\protect\astroncite{\emph{{Isella} et~al.}}{2012}]{isella:2012}
{Isella} A. et~al. (2012) \emph{\apj}, \emph{747}, 136.

\bibitem[\protect\astroncite{\emph{Jacquet}}{2013}]{Jacquet:2013p21188}
Jacquet E. (2013) \emph{Astronomy {\&} Astrophysics}, \emph{551}, 75.

\bibitem[\protect\astroncite{\emph{Johansen and
  Klahr}}{2005}]{Johansen:2005p8425}
Johansen A. and Klahr H. (2005) \emph{ApJ}, \emph{634}, 1353.

\bibitem[\protect\astroncite{\emph{{Johansen}
  et~al.}}{2009}]{2009ApJ...697.1269J}
{Johansen} A. et~al. (2009) \emph{\apj}, \emph{697}, 1269.

\bibitem[\protect\astroncite{\emph{{J{\o}rgensen}
  et~al.}}{2009}]{2009A&A...507..861J}
{J{\o}rgensen} J.~K. et~al. (2009) \emph{\aap}, \emph{507}, 861.

\bibitem[\protect\astroncite{\emph{{Juh{\'a}sz}
  et~al.}}{2010}]{2010ApJ...721..431J}
{Juh{\'a}sz} A. et~al. (2010) \emph{\apj}, \emph{721}, 431.

\bibitem[\protect\astroncite{\emph{{Juh{\'a}sz}
  et~al.}}{2012}]{2012ApJ...744..118J}
{Juh{\'a}sz} A. et~al. (2012) \emph{\apj}, \emph{744}, 118.

\bibitem[\protect\astroncite{\emph{{Juvela}
  et~al.}}{2013}]{2013A&A...556A..63J}
{Juvela} M. et~al. (2013) \emph{\aap}, \emph{556}, A63.

\bibitem[\protect\astroncite{\emph{{Keller} and
  {Gail}}}{2004}]{2004A&A...415.1177K}
{Keller} C. and {Gail} H.-P. (2004) \emph{\aap}, \emph{415}, 1177.

\bibitem[\protect\astroncite{\emph{{Kelly} et~al.}}{2012}]{2012ApJ...752...55K}
{Kelly} B.~C. et~al. (2012) \emph{\apj}, \emph{752}, 55.

\bibitem[\protect\astroncite{\emph{{Kempf} et~al.}}{1999}]{1999Icar..141..388K}
{Kempf} S. et~al. (1999) \emph{\icarus}, \emph{141}, 388.

\bibitem[\protect\astroncite{\emph{{Kessler-Silacci}
  et~al.}}{2006}]{2006ApJ...639..275K}
{Kessler-Silacci} J. et~al. (2006) \emph{\apj}, \emph{639}, 275.

\bibitem[\protect\astroncite{\emph{{Kessler-Silacci}
  et~al.}}{2007}]{2007ApJ...659..680K}
{Kessler-Silacci} J.~E. et~al. (2007) \emph{\apj}, \emph{659}, 680.

\bibitem[\protect\astroncite{\emph{{Keto} and
  {Caselli}}}{2008}]{2008ApJ...683..238K}
{Keto} E. and {Caselli} P. (2008) \emph{\apj}, \emph{683}, 238.

\bibitem[\protect\astroncite{\emph{{Klahr} and
  {Henning}}}{1997}]{1997Icar..128..213K}
{Klahr} H.~H. and {Henning} T. (1997) \emph{\icarus}, \emph{128}, 213.

\bibitem[\protect\astroncite{\emph{Kley and Lin}}{1992}]{Kley:1992p7134}
Kley W. and Lin D. N.~C. (1992) \emph{ApJ}, \emph{397}, 600.

\bibitem[\protect\astroncite{\emph{Kornet et~al.}}{2001}]{Kornet:2001p688}
Kornet K. et~al. (2001) \emph{A{\&}A}, \emph{378}, 180.

\bibitem[\protect\astroncite{\emph{{Kothe} et~al.}}{2010}]{2010ApJ...725.1242K}
{Kothe} S. et~al. (2010) \emph{\apj}, \emph{725}, 1242.

\bibitem[\protect\astroncite{\emph{{Kothe} et~al.}}{2013}]{2013Icar..225...75K}
{Kothe} S. et~al. (2013) \emph{\icarus}, \emph{225}, 75.

\bibitem[\protect\astroncite{\emph{{Krause} and
  {Blum}}}{2004}]{2004PhRvL..93b1103K}
{Krause} M. and {Blum} J. (2004) \emph{Physical Review Letters}, \emph{93}, 2,
  021103.

\bibitem[\protect\astroncite{\emph{Kretke and Lin}}{2007}]{Kretke:2007p697}
Kretke K.~A. and Lin D. N.~C. (2007) \emph{ApJ}, \emph{664}, L55.

\bibitem[\protect\astroncite{\emph{{Kruegel} and
  {Siebenmorgen}}}{1994}]{1994A&A...288..929K}
{Kruegel} E. and {Siebenmorgen} R. (1994) \emph{\aap}, \emph{288}, 929.

\bibitem[\protect\astroncite{\emph{{Kudo} et~al.}}{2008}]{2008ApJ...673L..67K}
{Kudo} T. et~al. (2008) \emph{\apjl}, \emph{673}, L67.

\bibitem[\protect\astroncite{\emph{{Kwon} et~al.}}{2009}]{2009ApJ...696..841K}
{Kwon} W. et~al. (2009) \emph{\apj}, \emph{696}, 841.

\bibitem[\protect\astroncite{\emph{{Laibe} et~al.}}{2008}]{2008A&A...487..265L}
{Laibe} G. et~al. (2008) \emph{\aap}, \emph{487}, 265.

\bibitem[\protect\astroncite{\emph{Laibe et~al.}}{2012}]{Laibe:2012p18591}
Laibe G. et~al. (2012) \emph{A{\&}A}, \emph{537}, 61.

\bibitem[\protect\astroncite{\emph{Lee}}{2000}]{Lee:2000p11145}
Lee M.~H. (2000) \emph{Icarus}, \emph{143}, 74.

\bibitem[\protect\astroncite{\emph{{Liu} et~al.}}{2012}]{2012A&A...546A...7L}
{Liu} Y. et~al. (2012) \emph{\aap}, \emph{546}, A7.

\bibitem[\protect\astroncite{\emph{Lodders}}{2003}]{Lodders:2003p6651}
Lodders K. (2003) \emph{ApJ}, \emph{591}, 1220.

\bibitem[\protect\astroncite{\emph{{Lommen}
  et~al.}}{2009}]{2009A&A...495..869L}
{Lommen} D. et~al. (2009) \emph{\aap}, \emph{495}, 869.

\bibitem[\protect\astroncite{\emph{{Luhman}}}{2007}]{2007ApJS..173..104L}
{Luhman} K.~L. (2007) \emph{\apjs}, \emph{173}, 104.

\bibitem[\protect\astroncite{\emph{{Luhman}
  et~al.}}{2010}]{2010ApJS..186..111L}
{Luhman} K.~L. et~al. (2010) \emph{\apjs}, \emph{186}, 111.

\bibitem[\protect\astroncite{\emph{Lynden-Bell and
  Pringle}}{1974}]{LyndenBell:1974p1945}
Lynden-Bell D. and Pringle J.~E. (1974) \emph{MNRAS}, \emph{168}, 603.

\bibitem[\protect\astroncite{\emph{{Mann} and
  {Williams}}}{2010}]{2010ApJ...725..430M}
{Mann} R.~K. and {Williams} J.~P. (2010) \emph{\apj}, \emph{725}, 430.

\bibitem[\protect\astroncite{\emph{Markiewicz
  et~al.}}{1991}]{Markiewicz:1991p9934}
Markiewicz W.~J. et~al. (1991) \emph{A{\&}A}, \emph{242}, 286.

\bibitem[\protect\astroncite{\emph{{Mathews}
  et~al.}}{2012}]{2012ApJ...745...23M}
{Mathews} G.~S. et~al. (2012) \emph{\apj}, \emph{745}, 23.

\bibitem[\protect\astroncite{\emph{{Mathews}
  et~al.}}{2013}]{2013A&A...557A.132M}
{Mathews} G.~S. et~al. (2013) \emph{\aap}, \emph{557}, A132.

\bibitem[\protect\astroncite{\emph{Mathis et~al.}}{1977}]{Mathis:1977p789}
Mathis J.~S. et~al. (1977) \emph{ApJ}, \emph{217}, 425.

\bibitem[\protect\astroncite{\emph{{Matthews}
  et~al.}}{2012}]{2012ApJ...744....8M}
{Matthews} L.~S. et~al. (2012) \emph{\apj}, \emph{744}, 8.

\bibitem[\protect\astroncite{\emph{{McClure}
  et~al.}}{2013}]{2013ApJ...775..114M}
{McClure} M.~K. et~al. (2013) \emph{\apj}, \emph{775}, 114.

\bibitem[\protect\astroncite{\emph{{Meru} et~al.}}{2013}]{2013ApJ...774L...4M}
{Meru} F. et~al. (2013) \emph{\apjl}, \emph{774}, L4.

\bibitem[\protect\astroncite{\emph{{Miyake} and
  {Nakagawa}}}{1993}]{1993Icar..106...20M}
{Miyake} K. and {Nakagawa} Y. (1993) \emph{\icarus}, \emph{106}, 20.

\bibitem[\protect\astroncite{\emph{Mizuno et~al.}}{1988}]{Mizuno:1988p20730}
Mizuno H. et~al. (1988) \emph{A{\&}A}, \emph{195}, 183.

\bibitem[\protect\astroncite{\emph{{Morfill} and
  {Voelk}}}{1984}]{1984ApJ...287..371M}
{Morfill} G.~E. and {Voelk} H.~J. (1984) \emph{\apj}, \emph{287}, 371.

\bibitem[\protect\astroncite{\emph{{Mulders} and
  {Dominik}}}{2012}]{2012A&A...539A...9M}
{Mulders} G.~D. and {Dominik} C. (2012) \emph{\aap}, \emph{539}, A9.

\bibitem[\protect\astroncite{\emph{{Mulders}
  et~al.}}{2013}]{2013A&A...549A.112M}
{Mulders} G.~D. et~al. (2013) \emph{\aap}, \emph{549}, A112.

\bibitem[\protect\astroncite{\emph{Nakagawa et~al.}}{1981}]{Nakagawa:1981p4533}
Nakagawa Y. et~al. (1981) \emph{Icarus}, \emph{45}, 517.

\bibitem[\protect\astroncite{\emph{{Nakagawa}
  et~al.}}{1986}]{1986Icar...67..375N}
{Nakagawa} Y. et~al. (1986) \emph{\icarus}, \emph{67}, 375.

\bibitem[\protect\astroncite{\emph{{Natta} and
  {Testi}}}{2004}]{2004ASPC..323..279N}
{Natta} A. and {Testi} L. (2004) in: \emph{Star Formation in the Interstellar
  Medium: In Honor of David Hollenbach}, vol. 323 of \emph{Astronomical Society
  of the Pacific Conference Series}, (edited by D.~{Johnstone}, F.~C. {Adams},
  D.~N.~C. {Lin}, D.~A. {Neufeeld}, and E.~C. {Ostriker}), pp. 279--+.

\bibitem[\protect\astroncite{\emph{{Natta} et~al.}}{2007}]{2007prpl.conf..767N}
{Natta} A. et~al. (2007) \emph{Protostars and Planets V}, pp. 767--781.

\bibitem[\protect\astroncite{\emph{Ohtsuki et~al.}}{1990}]{Ohtsuki:1990p799}
Ohtsuki K. et~al. (1990) \emph{Icarus}, \emph{83}, 205.

\bibitem[\protect\astroncite{\emph{Okuzumi}}{2009}]{Okuzumi:2009p7473}
Okuzumi S. (2009) \emph{ApJ}, \emph{698}, 1122.

\bibitem[\protect\astroncite{\emph{Okuzumi et~al.}}{2009}]{Okuzumi:2009p9772}
Okuzumi S. et~al. (2009) \emph{ApJ}, \emph{707}, 1247.

\bibitem[\protect\astroncite{\emph{Okuzumi et~al.}}{2011a}]{Okuzumi:2011p15377}
Okuzumi S. et~al. (2011a) \emph{ApJ}, \emph{731}, 95.

\bibitem[\protect\astroncite{\emph{{Okuzumi}
  et~al.}}{2011b}]{2011ApJ...731...96O}
{Okuzumi} S. et~al. (2011b) \emph{\apj}, \emph{731}, 96.

\bibitem[\protect\astroncite{\emph{{Okuzumi}
  et~al.}}{2012}]{2012ApJ...752..106O}
{Okuzumi} S. et~al. (2012) \emph{\apj}, \emph{752}, 106.

\bibitem[\protect\astroncite{\emph{{Oliveira}
  et~al.}}{2011}]{2011ApJ...734...51O}
{Oliveira} I. et~al. (2011) \emph{\apj}, \emph{734}, 51.

\bibitem[\protect\astroncite{\emph{{Ormel} and
  {Cuzzi}}}{2007}]{2007A&A...466..413O}
{Ormel} C.~W. and {Cuzzi} J.~N. (2007) \emph{\aap}, \emph{466}, 413.

\bibitem[\protect\astroncite{\emph{Ormel and Okuzumi}}{2013}]{Ormel:2013p21040}
Ormel C.~W. and Okuzumi S. (2013) \emph{ApJ}, \emph{771}, 44.

\bibitem[\protect\astroncite{\emph{{Ormel} et~al.}}{2007}]{2007A&A...461..215O}
{Ormel} C.~W. et~al. (2007) \emph{\aap}, \emph{461}, 215.

\bibitem[\protect\astroncite{\emph{{Ormel} et~al.}}{2008}]{2008ApJ...679.1588O}
{Ormel} C.~W. et~al. (2008) \emph{\apj}, \emph{679}, 1588.

\bibitem[\protect\astroncite{\emph{{Ormel} et~al.}}{2009}]{2009A&A...502..845O}
{Ormel} C.~W. et~al. (2009) \emph{\aap}, \emph{502}, 845.

\bibitem[\protect\astroncite{\emph{{Ormel} et~al.}}{2011}]{2011A&A...532A..43O}
{Ormel} C.~W. et~al. (2011) \emph{\aap}, \emph{532}, A43.

\bibitem[\protect\astroncite{\emph{{Ossenkopf}}}{1993}]{1993A&A...280..617O}
{Ossenkopf} V. (1993) \emph{\aap}, \emph{280}, 617.

\bibitem[\protect\astroncite{\emph{{Ossenkopf} and
  {Henning}}}{1994}]{1994A&A...291..943O}
{Ossenkopf} V. and {Henning} T. (1994) \emph{\aap}, \emph{291}, 943.

\bibitem[\protect\astroncite{\emph{{Pagani}
  et~al.}}{2010}]{2010Sci...329.1622P}
{Pagani} L. et~al. (2010) \emph{Science}, \emph{329}, 1622.

\bibitem[\protect\astroncite{\emph{Pan and Padoan}}{2010}]{Pan:2010p20918}
Pan L. and Padoan P. (2010) \emph{Journal of Fluid Mechanics}, \emph{661}, 73.

\bibitem[\protect\astroncite{\emph{{Pan} and
  {Padoan}}}{2013}]{2013ApJ...776...12P}
{Pan} L. and {Padoan} P. (2013) \emph{\apj}, \emph{776}, 12.

\bibitem[\protect\astroncite{\emph{{Pani{\'c}} et~al.}}{2008}]{panic:2008}
{Pani{\'c}} O. et~al. (2008) \emph{\aap}, \emph{491}, 219.

\bibitem[\protect\astroncite{\emph{{Paradis}
  et~al.}}{2010}]{2010A&A...520L...8P}
{Paradis} D. et~al. (2010) \emph{\aap}, \emph{520}, L8.

\bibitem[\protect\astroncite{\emph{{Paszun} and
  {Dominik}}}{2006}]{2006Icar..182..274P}
{Paszun} D. and {Dominik} C. (2006) \emph{\icarus}, \emph{182}, 274.

\bibitem[\protect\astroncite{\emph{{P{\'e}rez} et~al.}}{2012}]{perez:2012}
{P{\'e}rez} L.~M. et~al. (2012) \emph{\apjl}, \emph{760}, L17.

\bibitem[\protect\astroncite{\emph{{Pi{\'e}tu} et~al.}}{2007}]{pietu:2007}
{Pi{\'e}tu} V. et~al. (2007) \emph{\aap}, \emph{467}, 163.

\bibitem[\protect\astroncite{\emph{{Pinilla}
  et~al.}}{2012{\natexlab{a}}}]{2012A&A...545A..81P}
{Pinilla} P. et~al. (2012{\natexlab{a}}) \emph{\aap}, \emph{545}, A81.

\bibitem[\protect\astroncite{\emph{{Pinilla}
  et~al.}}{2012{\natexlab{b}}}]{2012A&A...538A.114P}
{Pinilla} P. et~al. (2012{\natexlab{b}}) \emph{\aap}, \emph{538}, A114.

\bibitem[\protect\astroncite{\emph{{Pinilla}
  et~al.}}{2013}]{2013A&A...554A..95P}
{Pinilla} P. et~al. (2013) \emph{\aap}, \emph{554}, A95.

\bibitem[\protect\astroncite{\emph{{Pinte} et~al.}}{2007}]{2007A&A...469..963P}
{Pinte} C. et~al. (2007) \emph{\aap}, \emph{469}, 963.

\bibitem[\protect\astroncite{\emph{{Pinte} et~al.}}{2008}]{2008A&A...489..633P}
{Pinte} C. et~al. (2008) \emph{\aap}, \emph{489}, 633.

\bibitem[\protect\astroncite{\emph{{Pollack}
  et~al.}}{1994}]{1994ApJ...421..615P}
{Pollack} J.~B. et~al. (1994) \emph{\apj}, \emph{421}, 615.

\bibitem[\protect\astroncite{\emph{{Poppe} et~al.}}{2000}]{2000ApJ...533..472P}
{Poppe} T. et~al. (2000) \emph{\apj}, \emph{533}, 472.

\bibitem[\protect\astroncite{\emph{{Qi} et~al.}}{2013}]{2013Sci...341..630Q}
{Qi} C. et~al. (2013) \emph{Science}, \emph{341}, 630.

\bibitem[\protect\astroncite{\emph{{Quanz} et~al.}}{2011}]{2011ApJ...738...23Q}
{Quanz} S.~P. et~al. (2011) \emph{\apj}, \emph{738}, 23.

\bibitem[\protect\astroncite{\emph{{Quanz} et~al.}}{2012}]{2012A&A...538A..92Q}
{Quanz} S.~P. et~al. (2012) \emph{\aap}, \emph{538}, A92.

\bibitem[\protect\astroncite{\emph{{Ricci}
  et~al.}}{2010{\natexlab{a}}}]{2010A&A...521A..66R}
{Ricci} L. et~al. (2010{\natexlab{a}}) \emph{\aap}, \emph{521}, A66.

\bibitem[\protect\astroncite{\emph{{Ricci}
  et~al.}}{2010{\natexlab{b}}}]{2010A&A...512A..15R}
{Ricci} L. et~al. (2010{\natexlab{b}}) \emph{\aap}, \emph{512}, A15.

\bibitem[\protect\astroncite{\emph{{Ricci} et~al.}}{2011}]{2011A&A...525A..81R}
{Ricci} L. et~al. (2011) \emph{\aap}, \emph{525}, A81.

\bibitem[\protect\astroncite{\emph{{Ricci}
  et~al.}}{2012{\natexlab{a}}}]{2012ApJ...761L..20R}
{Ricci} L. et~al. (2012{\natexlab{a}}) \emph{\apjl}, \emph{761}, L20.

\bibitem[\protect\astroncite{\emph{{Ricci}
  et~al.}}{2012{\natexlab{b}}}]{2012A&A...540A...6R}
{Ricci} L. et~al. (2012{\natexlab{b}}) \emph{\aap}, \emph{540}, A6.

\bibitem[\protect\astroncite{\emph{{Ricci} et~al.}}{2013}]{2013ApJ...764L..27R}
{Ricci} L. et~al. (2013) \emph{\apjl}, \emph{764}, L27.

\bibitem[\protect\astroncite{\emph{{Rodmann}
  et~al.}}{2006}]{2006A&A...446..211R}
{Rodmann} J. et~al. (2006) \emph{\aap}, \emph{446}, 211.

\bibitem[\protect\astroncite{\emph{Ros and Johansen}}{2013}]{Ros:2013p20760}
Ros K. and Johansen A. (2013) \emph{A{\&}A}, \emph{552}, 137.

\bibitem[\protect\astroncite{\emph{{Rosenfeld}
  et~al.}}{2013}]{2013ApJ...775..136R}
{Rosenfeld} K.~A. et~al. (2013) \emph{\apj}, \emph{775}, 136.

\bibitem[\protect\astroncite{\emph{{Roy} et~al.}}{2013}]{2013ApJ...763...55R}
{Roy} A. et~al. (2013) \emph{\apj}, \emph{763}, 55.

\bibitem[\protect\astroncite{\emph{Rozyczka
  et~al.}}{1994}]{Rozyczka:1994p20829}
Rozyczka M. et~al. (1994) \emph{Astrophysical Journal v.423}, \emph{423}, 736.

\bibitem[\protect\astroncite{\emph{{Sadavoy}
  et~al.}}{2013}]{2013ApJ...767..126S}
{Sadavoy} S.~I. et~al. (2013) \emph{\apj}, \emph{767}, 126.

\bibitem[\protect\astroncite{\emph{Safronov}}{1969}]{Safronov:1969p11177}
Safronov V.~S. (1969) \emph{Evolution of the protoplanetary cloud and formation
  of the earth and planets. English translation (1972)}.

\bibitem[\protect\astroncite{\emph{{Salter}
  et~al.}}{2010}]{2010A&A...521A..32S}
{Salter} D.~M. et~al. (2010) \emph{\aap}, \emph{521}, A32.

\bibitem[\protect\astroncite{\emph{{Schmitt}
  et~al.}}{1997}]{1997A&A...325..569S}
{Schmitt} W. et~al. (1997) \emph{\aap}, \emph{325}, 569.

\bibitem[\protect\astroncite{\emph{{Schr{\"a}pler} and
  {Blum}}}{2011}]{2011ApJ...734..108S}
{Schr{\"a}pler} R. and {Blum} J. (2011) \emph{\apj}, \emph{734}, 108.

\bibitem[\protect\astroncite{\emph{Schr{\"a}pler and
  Henning}}{2004}]{Schrapler:2004p2394}
Schr{\"a}pler R. and Henning T. (2004) \emph{ApJ}, \emph{614}, 960.

\bibitem[\protect\astroncite{\emph{{Scott} and
  {Krot}}}{2005}]{2005ApJ...623..571S}
{Scott} E.~R.~D. and {Krot} A.~N. (2005) \emph{\apj}, \emph{623}, 571.

\bibitem[\protect\astroncite{\emph{Shakura and
  Sunyaev}}{1973}]{Shakura:1973p4854}
Shakura N.~I. and Sunyaev R.~A. (1973) \emph{A{\&}A}, \emph{24}, 337.

\bibitem[\protect\astroncite{\emph{{Shetty}
  et~al.}}{2009}]{2009ApJ...696.2234S}
{Shetty} R. et~al. (2009) \emph{\apj}, \emph{696}, 2234.

\bibitem[\protect\astroncite{\emph{Shu et~al.}}{1994}]{Shu:1994p20679}
Shu F. et~al. (1994) \emph{ApJ}, \emph{429}, 781.

\bibitem[\protect\astroncite{\emph{Shu et~al.}}{2001}]{Shu:2001p20834}
Shu F.~H. et~al. (2001) \emph{ApJ}, \emph{548}, 1029.

\bibitem[\protect\astroncite{\emph{Sicilia-Aguilar
  et~al.}}{2007}]{SiciliaAguilar:2007p20899}
Sicilia-Aguilar A. et~al. (2007) \emph{ApJ}, \emph{659}, 1637.

\bibitem[\protect\astroncite{\emph{Sirono}}{2011{\natexlab{a}}}]{Sirono:2011p21038}
Sirono S.-I. (2011{\natexlab{a}}) \emph{ApJL}, \emph{733}, L41.

\bibitem[\protect\astroncite{\emph{Sirono}}{2011{\natexlab{b}}}]{Sirono:2011p21039}
Sirono S.-I. (2011{\natexlab{b}}) \emph{ApJ}, \emph{735}, 131.

\bibitem[\protect\astroncite{\emph{Smoluchowski}}{1916}]{Smoluchowski:1916p2203}
Smoluchowski M.~V. (1916) \emph{Physik. Zeit.}, \emph{17}, 557.

\bibitem[\protect\astroncite{\emph{{Steinacker}
  et~al.}}{2010}]{2010A&A...511A...9S}
{Steinacker} J. et~al. (2010) \emph{\aap}, \emph{511}, A9.

\bibitem[\protect\astroncite{\emph{{Stepinski} and
  {Valageas}}}{1996}]{1996A&A...309..301S}
{Stepinski} T.~F. and {Valageas} P. (1996) \emph{\aap}, \emph{309}, 301.

\bibitem[\protect\astroncite{\emph{{Stepinski} and
  {Valageas}}}{1997}]{1997A&A...319.1007S}
{Stepinski} T.~F. and {Valageas} P. (1997) \emph{\aap}, \emph{319}, 1007.

\bibitem[\protect\astroncite{\emph{{Stepnik}
  et~al.}}{2003}]{2003A&A...398..551S}
{Stepnik} B. et~al. (2003) \emph{\aap}, \emph{398}, 551.

\bibitem[\protect\astroncite{\emph{{Sterzik} and
  {Morfill}}}{1994}]{1994Icar..111..536S}
{Sterzik} M.~F. and {Morfill} G.~E. (1994) \emph{\icarus}, \emph{111}, 536.

\bibitem[\protect\astroncite{\emph{{Suttner} and
  {Yorke}}}{2001}]{2001ApJ...551..461S}
{Suttner} G. and {Yorke} H.~W. (2001) \emph{\apj}, \emph{551}, 461.

\bibitem[\protect\astroncite{\emph{{Suttner}
  et~al.}}{1999}]{1999ApJ...524..857S}
{Suttner} G. et~al. (1999) \emph{\apj}, \emph{524}, 857.

\bibitem[\protect\astroncite{\emph{{Suutarinen}
  et~al.}}{2013}]{2013A&A...555A.140S}
{Suutarinen} A. et~al. (2013) \emph{\aap}, \emph{555}, A140.

\bibitem[\protect\astroncite{\emph{Suyama et~al.}}{2012}]{Suyama:2012p20954}
Suyama T. et~al. (2012) \emph{ApJ}, \emph{753}, 115.

\bibitem[\protect\astroncite{\emph{Takeuchi and
  Lin}}{2002}]{Takeuchi:2002p3167}
Takeuchi T. and Lin D. N.~C. (2002) \emph{ApJ}, \emph{581}, 1344.

\bibitem[\protect\astroncite{\emph{{Tanaka}
  et~al.}}{2005}]{2005ApJ...625..414T}
{Tanaka} H. et~al. (2005) \emph{\apj}, \emph{625}, 414.

\bibitem[\protect\astroncite{\emph{{Teiser} and
  {Wurm}}}{2009{\natexlab{a}}}]{2009A&A...505..351T}
{Teiser} J. and {Wurm} G. (2009{\natexlab{a}}) \emph{\aap}, \emph{505}, 351.

\bibitem[\protect\astroncite{\emph{{Teiser} and
  {Wurm}}}{2009{\natexlab{b}}}]{2009MNRAS.393.1584T}
{Teiser} J. and {Wurm} G. (2009{\natexlab{b}}) \emph{\mnras}, \emph{393}, 1584.

\bibitem[\protect\astroncite{\emph{{Teiser}
  et~al.}}{2011}]{2011Icar..215..596T}
{Teiser} J. et~al. (2011) \emph{\icarus}, \emph{215}, 596.

\bibitem[\protect\astroncite{\emph{{Testi} et~al.}}{2001}]{2001ApJ...554.1087T}
{Testi} L. et~al. (2001) \emph{ApJ}, \emph{554}, 1087.

\bibitem[\protect\astroncite{\emph{{Testi} et~al.}}{2003}]{2003A&A...403..323T}
{Testi} L. et~al. (2003) \emph{A\&A}, \emph{403}, 323.

\bibitem[\protect\astroncite{\emph{{Tobin} et~al.}}{2013}]{2013ApJ...771...48T}
{Tobin} J.~J. et~al. (2013) \emph{\apj}, \emph{771}, 48.

\bibitem[\protect\astroncite{\emph{{Trotta}
  et~al.}}{2013}]{2013A&A...558A..64T}
{Trotta} F. et~al. (2013) \emph{\aap}, \emph{558}, A64.

\bibitem[\protect\astroncite{\emph{{Ubach} et~al.}}{2012}]{2012MNRAS.425.3137U}
{Ubach} C. et~al. (2012) \emph{\mnras}, \emph{425}, 3137.

\bibitem[\protect\astroncite{\emph{{Uribe} et~al.}}{2011}]{2011ApJ...736...85U}
{Uribe} A.~L. et~al. (2011) \emph{\apj}, \emph{736}, 85.

\bibitem[\protect\astroncite{\emph{{Urpin}}}{1984}]{1984SvA....28...50U}
{Urpin} V.~A. (1984) \emph{\sovast}, \emph{28}, 50.

\bibitem[\protect\astroncite{\emph{{van Boekel}
  et~al.}}{2004}]{2004Natur.432..479V}
{van Boekel} R. et~al. (2004) \emph{\nat}, \emph{432}, 479.

\bibitem[\protect\astroncite{\emph{{van Boekel}
  et~al.}}{2005}]{2005A&A...437..189V}
{van Boekel} R. et~al. (2005) \emph{\aap}, \emph{437}, 189.

\bibitem[\protect\astroncite{\emph{{van der Marel}
  et~al.}}{2013}]{2013Sci...340.1199V}
{van der Marel} N. et~al. (2013) \emph{Science}, \emph{340}, 1199.

\bibitem[\protect\astroncite{\emph{{Veneziani}
  et~al.}}{2010}]{2010ApJ...713..959V}
{Veneziani} M. et~al. (2010) \emph{\apj}, \emph{713}, 959.

\bibitem[\protect\astroncite{\emph{{Voelk} et~al.}}{1980}]{1980A&A....85..316V}
{Voelk} H.~J. et~al. (1980) \emph{\aap}, \emph{85}, 316.

\bibitem[\protect\astroncite{\emph{{Wada} et~al.}}{2008}]{2008ApJ...677.1296W}
{Wada} K. et~al. (2008) \emph{\apj}, \emph{677}, 1296.

\bibitem[\protect\astroncite{\emph{{Wada} et~al.}}{2009}]{2009ApJ...702.1490W}
{Wada} K. et~al. (2009) \emph{\apj}, \emph{702}, 1490.

\bibitem[\protect\astroncite{\emph{{Weidenschilling}}}{1977}]{1977MNRAS.180...57W}
{Weidenschilling} S.~J. (1977) \emph{\mnras}, \emph{180}, 57.

\bibitem[\protect\astroncite{\emph{Weidenschilling}}{1977}]{Weidenschilling:1977p15694}
Weidenschilling S.~J. (1977) \emph{Ap{\&}SS}, \emph{51}, 153.

\bibitem[\protect\astroncite{\emph{{Weidenschilling}}}{1980}]{1980Icar...44..172W}
{Weidenschilling} S.~J. (1980) \emph{\icarus}, \emph{44}, 172.

\bibitem[\protect\astroncite{\emph{{Weidenschilling}}}{1984}]{1984Icar...60..553W}
{Weidenschilling} S.~J. (1984) \emph{\icarus}, \emph{60}, 553.

\bibitem[\protect\astroncite{\emph{Weidenschilling}}{1997}]{Weidenschilling:1997p4593}
Weidenschilling S.~J. (1997) \emph{Icarus}, \emph{127}, 290.

\bibitem[\protect\astroncite{\emph{{Weidenschilling} and
  {Cuzzi}}}{1993}]{1993prpl.conf.1031W}
{Weidenschilling} S.~J. and {Cuzzi} J.~N. (1993) in: \emph{Protostars and
  Planets III}, (edited by E.~H. {Levy} and J.~I. {Lunine}), pp. 1031--1060.

\bibitem[\protect\astroncite{\emph{{Weidenschilling} and
  {Ruzmaikina}}}{1994}]{1994ApJ...430..713W}
{Weidenschilling} S.~J. and {Ruzmaikina} T.~V. (1994) \emph{\apj}, \emph{430},
  713.

\bibitem[\protect\astroncite{\emph{{Weintraub}
  et~al.}}{1989}]{1989ApJ...340L..69W}
{Weintraub} D.~A. et~al. (1989) \emph{\apjl}, \emph{340}, L69.

\bibitem[\protect\astroncite{\emph{{Whipple}}}{1972}]{1972fpp..conf..211W}
{Whipple} F.~L. (1972) in: \emph{From Plasma to Planet}, (edited by
  A.~{Elvius}), p. 211.

\bibitem[\protect\astroncite{\emph{Whipple}}{1972}]{Whipple:1972p4621}
Whipple F.~L. (1972) \emph{From Plasma to Planet}, p. 211.

\bibitem[\protect\astroncite{\emph{{Williams} and
  {Cieza}}}{2011}]{2011ARA&A..49...67W}
{Williams} J.~P. and {Cieza} L.~A. (2011) \emph{\araa}, \emph{49}, 67.

\bibitem[\protect\astroncite{\emph{{Williams}
  et~al.}}{2005}]{2005ApJ...634..495W}
{Williams} J.~P. et~al. (2005) \emph{\apj}, \emph{634}, 495.

\bibitem[\protect\astroncite{\emph{{Williams}
  et~al.}}{2013}]{2013MNRAS.435.1671W}
{Williams} J.~P. et~al. (2013) \emph{\mnras}, \emph{435}, 1671.

\bibitem[\protect\astroncite{\emph{{Wilner}
  et~al.}}{2005}]{2005ApJ...626L.109W}
{Wilner} D.~J. et~al. (2005) \emph{ApJL}, \emph{626}, L109.

\bibitem[\protect\astroncite{\emph{{Windmark}
  et~al.}}{2012{\natexlab{a}}}]{2012A&A...544L..16W}
{Windmark} F. et~al. (2012{\natexlab{a}}) \emph{\aap}, \emph{544}, L16.

\bibitem[\protect\astroncite{\emph{{Windmark}
  et~al.}}{2012{\natexlab{b}}}]{2012A&A...540A..73W}
{Windmark} F. et~al. (2012{\natexlab{b}}) \emph{\aap}, \emph{540}, A73.

\bibitem[\protect\astroncite{\emph{{Wisniewski}
  et~al.}}{2008}]{2008ApJ...682..548W}
{Wisniewski} J.~P. et~al. (2008) \emph{\apj}, \emph{682}, 548.

\bibitem[\protect\astroncite{\emph{{Wolf} et~al.}}{2003}]{2003ApJ...588..373W}
{Wolf} S. et~al. (2003) \emph{\apj}, \emph{588}, 373.

\bibitem[\protect\astroncite{\emph{{Wolf} et~al.}}{2008}]{2008ApJ...674L.101W}
{Wolf} S. et~al. (2008) \emph{\apjl}, \emph{674}, L101.

\bibitem[\protect\astroncite{\emph{{Wolf} et~al.}}{2012}]{2012A&ARv..20...52W}
{Wolf} S. et~al. (2012) \emph{\aapr}, \emph{20}, 52.

\bibitem[\protect\astroncite{\emph{{Woody} et~al.}}{1989}]{1989ApJ...337L..41W}
{Woody} D.~P. et~al. (1989) \emph{\apjl}, \emph{337}, L41.

\bibitem[\protect\astroncite{\emph{{Wurm} et~al.}}{2005}]{2005Icar..178..253W}
{Wurm} G. et~al. (2005) \emph{\icarus}, \emph{178}, 253.

\bibitem[\protect\astroncite{\emph{Youdin and
  Goodman}}{2005}]{Youdin:2005p11585}
Youdin A.~N. and Goodman J. (2005) \emph{ApJ}, \emph{620}, 459.

\bibitem[\protect\astroncite{\emph{Youdin and
  Lithwick}}{2007}]{Youdin:2007p2021}
Youdin A.~N. and Lithwick Y. (2007) \emph{Icarus}, \emph{192}, 588.

\bibitem[\protect\astroncite{\emph{{Youdin} and
  {Shu}}}{2002}]{2002ApJ...580..494Y}
{Youdin} A.~N. and {Shu} F.~H. (2002) \emph{\apj}, \emph{580}, 494.

\bibitem[\protect\astroncite{\emph{Zhukovska
  et~al.}}{2008}]{Zhukovska:2008p16575}
Zhukovska S. et~al. (2008) \emph{A{\&}A}, \emph{479}, 453.

\bibitem[\protect\astroncite{\emph{{Zsom} et~al.}}{2010}]{2010A&A...513A..57Z}
{Zsom} A. et~al. (2010) \emph{\aap}, \emph{513}, A57.

\bibitem[\protect\astroncite{\emph{Zsom et~al.}}{2011}]{Zsom:2011p21105}
Zsom A. et~al. (2011) \emph{\aap}, \emph{534}, 73.

\end{thebibliography}

\end{document}